\documentclass[ALICE,manyauthors]{cernphprep}
\usepackage[comma,square,numbers,sort&compress]{natbib}

\usepackage{lineno}
\usepackage{xspace}
\usepackage{hyperref}
\usepackage{siunitx}
\usepackage[T1]{fontenc}
\usepackage{xcolor}
\usepackage{multirow}
\usepackage{makecell}
\usepackage{booktabs}
\usepackage{upgreek}
\usepackage{comment}
\usepackage{amsmath}
\usepackage{xfrac}
\usepackage{amssymb}
\usepackage{orcidlink}

\begin{document}
%

\newcommand{\pp}           {pp\xspace}

\newcommand{\roots}        {\ensuremath{\sqrt{s}}\xspace}
\newcommand{\pt}           {\ensuremath{p_{\rm T}}\xspace}
\newcommand{\pT}           {\pt}
\newcommand{\px}           {\ensuremath{p_{\rm x}}\xspace}
\newcommand{\py}           {\ensuremath{p_{\rm y}}\xspace}
\newcommand{\etarange}[1]  {\mbox{$\left| \eta \right| < #1$}}
\newcommand{\yrange}[1]    {\mbox{$\left| y \right| < #1$}}
\newcommand{\dndy}         {\ensuremath{\mathrm{d}N_\mathrm{ch}/\mathrm{d}y}\xspace}
\newcommand{\dndeta}       {\ensuremath{\mathrm{d}N_\mathrm{ch}/\mathrm{d}\eta}\xspace}
\newcommand{\avdndeta}     {\ensuremath{\langle\dndeta\rangle}\xspace}
\newcommand{\dNdy}         {\ensuremath{\mathrm{d}N_\mathrm{ch}/\mathrm{d}y}\xspace}
\newcommand{\dEdx}         {\ensuremath{\textrm{d}E/\textrm{d}x}\xspace}
\newcommand{\kstar}        {\ensuremath{k^*}\xspace}
\newcommand{\rstar}     {\ensuremath{r^{*}}\xspace}

\newcommand{\mt}           {\ensuremath{m_{\rm{T}}}\xspace}
\newcommand{\St}           {\ensuremath{S_{\rm{T}}}\xspace}

\newcommand{\seven}        {$\sqrt{s} =7$~Te\kern-.1emV\xspace}
\newcommand{\thirteen}     {$\sqrt{s} =13$~Te\kern-.1emV\xspace}
\newcommand{\five}         {$\sqrt{s} =5.02$~Te\kern-.1emV\xspace}
\newcommand{\tev}          {Te\kern-.1emV\xspace}
\newcommand{\lumi}         {\ensuremath{\mathcal{L}}\xspace}
\newcommand{\mum}          {\ensuremath{\upmu\mathrm{m}}\xspace}
\newcommand{\fm}           {\ensuremath{\mathrm{fm}}\xspace}
\newcommand{\centm}        {\ensuremath{\mathrm{cm}}\xspace}

\newcommand{\ITS}          {\ensuremath{\mathrm{ITS}}\xspace}
\newcommand{\TOF}          {\ensuremath{\mathrm{TOF}}\xspace}
\newcommand{\ZDC}          {\ensuremath{\mathrm{ZDC}}\xspace}
\newcommand{\ZDCs}         {\ensuremath{\mathrm{ZDCs}}\xspace}
\newcommand{\ZNA}          {\ensuremath{\mathrm{ZNA}}\xspace}
\newcommand{\ZNC}          {\ensuremath{\mathrm{ZNC}}\xspace}
\newcommand{\SPD}          {\ensuremath{\mathrm{SPD}}\xspace}
\newcommand{\SDD}          {\ensuremath{\mathrm{SDD}}\xspace}
\newcommand{\SSD}          {\ensuremath{\mathrm{SSD}}\xspace}
\newcommand{\TPC}          {\ensuremath{\mathrm{TPC}}\xspace}
\newcommand{\TRD}          {\ensuremath{\mathrm{TRD}}\xspace}
\newcommand{\VZERO}        {\ensuremath{\mathrm{V0}}\xspace}
\newcommand{\VZEROA}       {\ensuremath{\mathrm{V0A}}\xspace}
\newcommand{\VZEROC}       {\ensuremath{\mathrm{V0C}}\xspace}
\newcommand{\Vdecay} 	   {\ensuremath{\mathrm{V}^0}\xspace}

\newcommand{\ee}           {\ensuremath{\mathrm{e}^{+}\mathrm{e}^{-}}} 
\newcommand{\pip}          {\ensuremath{\uppi^{+}}\xspace}
\newcommand{\pim}          {\ensuremath{\uppi^{-}}\xspace}
\newcommand{\kap}          {\ensuremath{\mathrm{K}^{+}}\xspace}
\newcommand{\kam}          {\ensuremath{\mathrm{K}^{-}}\xspace}
\newcommand{\KK}           {\ensuremath{\mathrm{K}^{+}\mathrm{K}^{-}}\xspace}
\newcommand{\KKbar}        {\ensuremath{\mathrm{K}\overline{\mathrm{K}}}\xspace}
\newcommand{\phiPart}      {\ensuremath{\phi}\xspace}
\newcommand{\pbar}         {\ensuremath{\mathrm\overline{p}}\xspace}
\newcommand{\kzero}        {\ensuremath{{\mathrm K}^{0}_{\mathrm{S}}}\xspace}
\newcommand{\pipm}         {\ensuremath{\uppi^\pm}\xspace}
\newcommand{\Kpm}         {\ensuremath{\mathrm{K}^\pm}\xspace}
\newcommand{\lmb}          {\ensuremath{\Lambda}\xspace}
\newcommand{\almb}         {\ensuremath{\overline{\Lambda}}\xspace}
\newcommand{\Om}           {\ensuremath{\Omega^-}\xspace}
\newcommand{\Mo}           {\ensuremath{\overline{\Omega}^+}\xspace}
\newcommand{\X}            {\ensuremath{\Xi^-}\xspace}
\newcommand{\Ix}           {\ensuremath{\overline{\Xi}^+}\xspace}
\newcommand{\Xis}          {\ensuremath{\Xi^{\pm}}\xspace}
\newcommand{\Oms}          {\ensuremath{\Omega^{\pm}}\xspace}

\newcommand{\Led}          {Lednick\'y--Lyuboshits\xspace}
\newcommand{\Ledn}         {Lednick\'y--Lyuboshits approach\xspace}

\newcommand{\pP}           {\ensuremath{\mathrm{p}\mbox{--}\mathrm{p}}\xspace}
\newcommand{\ApAP}         {\ensuremath{\pbar\mbox{--}\pbar}\xspace}
\newcommand{\pPComb}       {\ensuremath{\mathrm{p}\mbox{--}\mathrm{p} \oplus \pbar\mbox{--}\pbar}\xspace}

\newcommand{\Dminus}                 {\ensuremath{\mathrm{D^-}}\xspace}
\newcommand{\Dplus}                  {\ensuremath{\mathrm{D^+}}\xspace}
\newcommand{\Dstar}                  {\ensuremath{\mathrm{D^*}}\xspace}
\newcommand{\Dstarp}                 {\ensuremath{\mathrm{D^{*+}}}\xspace}
\newcommand{\Dstarm}                 {\ensuremath{\mathrm{D^{*-}}}\xspace}
\newcommand{\Dstarpm}                {\ensuremath{\mathrm{D^{*\pm}}}\xspace}
\newcommand{\Dzero}                  {\ensuremath{\mathrm{D^0}}\xspace}
\newcommand{\Dpm}                    {\ensuremath{\mathrm{D}^\pm}\xspace}
\newcommand{\DorDstar}               {\ensuremath{\mathrm{D^{(*)\pm}}}\xspace}
\newcommand{\DorDstarp}              {\ensuremath{\mathrm{D^{(*)+}}}\xspace}
\newcommand{\DorDstarpm}             {\ensuremath{\mathrm{D^{(*)\pm}}}\xspace}
\newcommand{\DorDstarNoSign}         {\ensuremath{\mathrm{D^{(*)}}}\xspace}
\newcommand{\DorDstarNoSignPi}       {\ensuremath{\mathrm{D^{(*)}}\uppi}\xspace}
\newcommand{\DorDstarNoSignK}        {\ensuremath{\mathrm{D^{(*)}K}}\xspace}
\newcommand{\Bplusminus}             {\ensuremath{\mathrm{B}^{\pm}}\xspace}
\newcommand{\Tcc}                    {\ensuremath{\mathrm{T_{cc}^+}}\xspace}
\newcommand{\Chic}                   {\ensuremath{\upchi_\mathrm{c1}(3872)}\xspace}
\newcommand{\Pc}[1]                  {\ensuremath{\mathrm{P_\mathrm{c}(#1)}}\xspace} 
\newcommand{\Dsexc}[2]               {\ensuremath{\mathrm{D_{s#1}^*}(#2)}\xspace}
\newcommand{\DsexcNostar}[2]         {\ensuremath{\mathrm{D_{s#1}}(#2)}\xspace}
\newcommand{\Dexc}[2]                {\ensuremath{\mathrm{D_{#1}^*}(#2)}\xspace}
\newcommand{\DexcNostar}[2]          {\ensuremath{\mathrm{D_{#1}}(#2)}\xspace}

\newcommand{\nDbar}        {\ensuremath{\mbox{n}\overline{\mathrm{D}^{0}}}\xspace}
\newcommand{\pD}           {\ensuremath{\mbox{pD}^{-}}\xspace}
\newcommand{\KpDp}         {\ensuremath{\mbox{K}^{+}\mbox{D}^{+}}\xspace}
\newcommand{\KmDm}         {\ensuremath{\mbox{K}^{-}\mbox{D}^{-}}\xspace}
\newcommand{\KpDm}         {\ensuremath{\mbox{K}^{+}\mbox{D}^{-}}\xspace}
\newcommand{\KmDp}         {\ensuremath{\mbox{K}^{-}\mbox{D}^{+}}\xspace}
\newcommand{\pipDp}        {\ensuremath{\uppi^+\mbox{D}^+}\xspace}
\newcommand{\pimDm}        {\ensuremath{\uppi^-\mbox{D}^-}\xspace}
\newcommand{\pipDm}        {\ensuremath{\uppi^+\mbox{D}^-}\xspace}
\newcommand{\pimDp}        {\ensuremath{\uppi^-\mbox{D}^+}\xspace}

\newcommand{\apaD} {\ensuremath{\overline{\mathrm{p}}\mbox{D}^{+}}\xspace}
\newcommand{\pDcomb} {\ensuremath{\mbox{pD}^{-} \oplus \overline{\mathrm{p}}\mbox{D}^{+}}\xspace}
\newcommand{\piD} {\ensuremath{\uppi^\pm\mbox{D}^{-}}\xspace}
\newcommand{\KD} {\ensuremath{\mbox{K}^\pm\mbox{D}^{-}}\xspace}

\newcommand{\DPi}     {\ensuremath{\mathrm{D}\uppi}\xspace}
\newcommand{\DK}      {\ensuremath{\mathrm{DK}}\xspace}
\newcommand{\Deta}    {\ensuremath{\mathrm{D\upeta}}\xspace}
\newcommand{\DsK}     {\ensuremath{\mathrm{D_{s}K}}\xspace}
\newcommand{\DsAntiK} {\ensuremath{\mathrm{D_{s}\overline{K}}}\xspace}
\newcommand{\DAntiK} {\ensuremath{\mathrm{D\overline{K}}}\xspace}
\newcommand{\DstarAntiK} {\ensuremath{\mathrm{D^*\overline{K}}}\xspace}

\newcommand{\pKpipi}       {\ensuremath{\mbox{p(K}^{+}\uppi^{-}\uppi^{-}\mathrm{)}}\xspace}
\newcommand{\Kpipi}        {\ensuremath{\mbox{(K}^{+}\uppi^{-}\uppi^{-}\mathrm{)}}\xspace}
\newcommand{\piDSC}        {\ensuremath{\pipDp \oplus \pimDm}\xspace}
\newcommand{\piDOC}        {\ensuremath{\pipDm \oplus \pimDp}\xspace}
\newcommand{\KDSC}         {\ensuremath{\KpDp \oplus \KmDm }\xspace}
\newcommand{\KDOC}         {\ensuremath{\KpDm \oplus \KmDp }\xspace}

\newcommand{\DPipp} {\ensuremath{\mathrm{D^{+}\uppi^+}}\xspace}
\newcommand{\DPimm} {\ensuremath{\mathrm{D^{-}\uppi^-}}\xspace}
\newcommand{\DPipm} {\ensuremath{\mathrm{D^{+}\uppi^-}}\xspace}
\newcommand{\DPimp} {\ensuremath{\mathrm{D^{-}\uppi^+}}\xspace}
\newcommand{\DPisc} {\ensuremath{\DPipp \oplus \DPimm}\xspace}
\newcommand{\DPioc} {\ensuremath{\DPipm \oplus \DPimp}\xspace}

\newcommand{\DKpp} {\ensuremath{\mathrm{D^{+}\mathrm{K}^+}}\xspace}
\newcommand{\DKmm} {\ensuremath{\mathrm{D^{-}\mathrm{K}^-}}\xspace}
\newcommand{\DKpm} {\ensuremath{\mathrm{D^{+}\mathrm{K}^-}}\xspace}
\newcommand{\DKmp} {\ensuremath{\mathrm{D^{-}\mathrm{K}^+}}\xspace}
\newcommand{\DKsc} {\ensuremath{\DKpp \oplus \DKmm}\xspace}
\newcommand{\DKoc} {\ensuremath{\DKpm \oplus \DKmp}\xspace}

\newcommand{\DstarPipp} {\ensuremath{\mathrm{D^{*+}\uppi^+}}\xspace}
\newcommand{\DstarPimm} {\ensuremath{\mathrm{D^{*-}\uppi^-}}\xspace}
\newcommand{\DstarPipm} {\ensuremath{\mathrm{D^{*+}\uppi^-}}\xspace}
\newcommand{\DstarPimp} {\ensuremath{\mathrm{D^{*-}\uppi^+}}\xspace}
\newcommand{\DstarPisc} {\ensuremath{\DstarPipp \oplus \DstarPimm}\xspace}
\newcommand{\DstarPioc} {\ensuremath{\DstarPipm \oplus \DstarPimp}\xspace}

\newcommand{\DstarKpp} {\ensuremath{\mathrm{D^{*+}K^+}}\xspace}
\newcommand{\DstarKmm} {\ensuremath{\mathrm{D^{*-}K^-}}\xspace}
\newcommand{\DstarKpm} {\ensuremath{\mathrm{D^{*+}K^-}}\xspace}
\newcommand{\DstarKmp} {\ensuremath{\mathrm{D^{*-}K^+}}\xspace}
\newcommand{\DstarKsc} {\ensuremath{\DstarKpp \oplus \DstarKmm}\xspace}
\newcommand{\DstarKoc} {\ensuremath{\DstarKpm \oplus \DstarKpm}\xspace}

\newcommand{\DDstarbar}  {\ensuremath{\mathrm{D\overline{D}^*}}\xspace}
\newcommand{\DDstar}     {\ensuremath{\mathrm{DD^*}}\xspace}
\newcommand{\SigmacDbar} {\ensuremath{\Sigma_\mathrm{c}\mathrm{\overline{D}}}\xspace}
\newcommand{\SigmacDstarbar} {\ensuremath{\Sigma_\mathrm{c}\mathrm{\overline{D}^*}}\xspace}

\newcommand{\Lednicky} {Lednick\'y--Lyuboshits\xspace}

\newcommand{\da}{\partial}
\newcommand{\de}{\mathrm{d}}
\newcommand{\temp}[1]{\textbf{\textcolor{red}{#1}}}
\newcommand{\dpl}{\rm{D}^+}
\newcommand{\dst}{\rm{D}^{*+}}
\newcommand{\av}[1]{\left\langle #1 \right\rangle}

\newcommand{\sqrts}{\sqrt{s}}
\newcommand{\sqrtsNN}{\sqrt{s_{\scriptscriptstyle \rm NN}}}
\newcommand{\pPb}{\mbox{p--Pb}}
\newcommand{\GeVc}{\mathrm{GeV}/c}
\newcommand{\GeVcc}{\mathrm{GeV}/c^2}
\newcommand{\MeVc}{\mathrm{MeV}/c}
\newcommand{\MeVcc}{\mathrm{MeV}/c^2}
\newcommand{\mb}{\mathrm{mb}}
\newcommand{\mub}{\upmu\mathrm{b}}

\newcommand{\DzerotoKpi}       {\ensuremath{\mathrm{D^0 \to K^-\uppi^+}}\xspace}
\newcommand{\DplustoKpipi}     {\ensuremath{\mathrm{D^+\to K^-\uppi^+\uppi^+}}\xspace}
\newcommand{\DstartoDpi}       {\ensuremath{\mathrm{D^{*+} \to D^0 \uppi^+}}\xspace}
\newcommand{\DstartoKpipi}     {\ensuremath{\mathrm{D^{*+} \to D^0 \uppi^+ \to K^-\uppi^+\uppi^+}}\xspace}
\newcommand{\Dstophipi}        {\ensuremath{\mathrm{D_s^+\to \upphi\uppi^+}}\xspace}
\newcommand{\Dstophipipm}      {\ensuremath{\mathrm{D_s^\pm\to \upphi\uppi^\pm}}\xspace}
\newcommand{\DstophipitoKKpi}  {\ensuremath{\mathrm{D_s^+\to \upphi\uppi^+\to K^-K^+\uppi^+}}\xspace}
\newcommand{\DplustoKKpi}      {\ensuremath{\mathrm{D^+\to K^-K^+\uppi^+}}\xspace}
\newcommand{\DstarPi}{\ensuremath{\Dstar\uppi}\xspace}
\newcommand{\DstarK} {\ensuremath{\Dstar\mathrm{K}}\xspace}

\newcommand{\fReal} {\ensuremath{\Re(f_0) =\nolinebreak 0.85 \pm\nolinebreak 0.34\,(\mathrm{stat.}) \pm\nolinebreak 0.14\,(\mathrm{syst.})~\si{fm}}\xspace}
\newcommand{\fImag} {\ensuremath{\Im(f_0) =\nolinebreak 0.16 \pm\nolinebreak 0.10\,(\mathrm{stat.}) \pm\nolinebreak 0.09\,(\mathrm{syst.})~\si{fm}}\xspace}
\newcommand{\dZero} {\ensuremath{d_0 =\nolinebreak 7.85 \pm\nolinebreak 1.54\,(\mathrm{stat.})  \pm\nolinebreak 0.26\,(\mathrm{syst.})~\si{fm}}\xspace}

\newcommand{\NsigmaNoCorr} {\ensuremath{5.7 \pm\nolinebreak 0.8\,(\mathrm{stat.})  \pm\nolinebreak 0.5\,(\mathrm{syst.})~\sigma}\xspace }

\newcommand{\NsigmaNoCorrRange} {\ensuremath{4.7 - 6.6~\sigma}\xspace}

\newcommand{\VeffGauss} {\ensuremath{V_{\mathrm{eff}}=\nolinebreak 2.5\pm\nolinebreak 0.9 \,(\mathrm{stat.})\pm\nolinebreak 1.4\,(\mathrm{syst.})~\si{\MeV}}\xspace}
\newcommand{\muGauss}   {\ensuremath{\upmu=\nolinebreak0.14\pm\nolinebreak 0.06 \,(\mathrm{stat.})\pm\nolinebreak 0.09\,(\mathrm{syst.})~\si{fm^{-2}}}\xspace}

\newcommand{\alphaYuk}  {\ensuremath{\upalpha=\nolinebreak65.9 \pm\nolinebreak 38.0\,(\mathrm{stat.})\pm\nolinebreak 17.5\,(\mathrm{syst.})~\si{\MeV}}\xspace}
\newcommand{\AYuk}  {\ensuremath{A=\nolinebreak0.021\pm\nolinebreak 0.009\,(\mathrm{stat.})\pm\nolinebreak 0.006\,(\mathrm{syst.})}\xspace}
\newcommand{\gYuk}  {\ensuremath{g_{\Nphi} =\nolinebreak 0.14\pm\nolinebreak 0.03\,(\mathrm{stat.})\pm\nolinebreak 0.02\,(\mathrm{syst.})}\xspace}

\newcommand{\norm} {\ensuremath{\mathit{M} = 0.96}\xspace}

\newcommand{\realScatSigma}{\ensuremath{2.3\upsigma}\xspace}

\newcommand{\pythia} {\textsc{Pythia}\xspace}
\newcommand{\geant} {\textsc{Geant}\xspace}
\newcommand{\epos} {\textsc{Epos}\xspace}
\newcommand{\fist}{\textsc{ThermalFIST}\xspace}
\newcommand{\hipeforml} {hipe4ML\xspace}

\hyphenation{ALICE}

\begin{titlepage}
\PHyear{2024}       
\PHnumber{013}      
\PHdate{21 January}  

\title{Studying the interaction between charm and light-flavor mesons}
\ShortTitle{Studying the interaction between charm and light-flavor mesons}   

\Collaboration{ALICE Collaboration\thanks{See Appendix~\ref{app:collab} for the list of collaboration members}}
\ShortAuthor{ALICE Collaboration} 

\begin{abstract}
    The two-particle momentum correlation functions between charm mesons (\Dstarpm and \Dpm) and charged light-flavor mesons ($\uppi^{\pm}$ and K$^{\pm}$) in all charge-combinations are measured for the first time by the ALICE Collaboration in high-multiplicity
    proton--proton collisions at a center-of-mass energy of \thirteen. 
    For \DK and \DstarK pairs, the experimental results are in agreement with theoretical predictions of the residual strong interaction based on quantum chromodynamics calculations on the lattice and chiral effective field theory. In the case of \DPi and \DstarPi pairs, tension between the calculations including strong interactions and the measurement is observed. For all particle pairs, the data can be adequately described by Coulomb interaction only, indicating a shallow interaction between charm and light-flavor mesons.
    Finally, the scattering lengths governing the residual strong interaction of the \DPi and \DstarPi systems are determined by fitting the experimental correlation functions with a model that employs a Gaussian potential. The extracted values are small and compatible with zero.
    \end{abstract}
\end{titlepage}

\setcounter{page}{2} 


\section{Introduction}

The exploration of the strong interaction within hadrons remains a pivotal question in particle physics. Quantum chromodynamics (QCD) has been well tested at distances significantly shorter than the nucleon's size and many high-energy phenomena can be effectively explained through perturbative QCD at the quark level. However, when the distance between quarks reaches the nucleon size, the QCD becomes a strongly coupled theory and the low-energy processes between hadrons are not yet well described.
From the experimental point of view, the residual strong interaction between hadrons has been studied in the past using scattering experiments at low energies with both stable and unstable beams. Numerous results have been achieved for nucleon--nucleon interactions with this method~\cite{Arndt:2007qn,NavarroPerez:2013usk}, however, due to the experimental challenge in realizing scattering experiments with unstable particles, only a reduced set of measurements could have been performed in the strange sector and none in the charm sector.
In order to overcome these experimental limitations, the femtoscopy technique has emerged as an interesting tool to study
reactions among hadrons~\cite{Pratt:1986cc}.
This method is based on the measurement of the correlation function  of pairs of hadrons in momentum space, which encodes the information of the interaction between the two hadrons convoluted with the emitting source distribution. The ALICE Collaboration measured the residual strong interaction between several light and strange hadrons using the femtoscopy technique in high-multiplicity proton--proton (pp) collisions, including pp, pK$^{\pm}$, p$\Lambda$, p$\overline{\Lambda}$, p$\Sigma^0$, $\Lambda\Lambda$, $\Lambda\overline{\Lambda}$,  p$\Xi^-$, p$\Omega^-$, p$\upphi$, and $\Lambda$K interactions~\cite{ALICE:2018ysd,ALICE:2019gcn,ALICE:2021njx,ALICE:2019buq,ALICE:2019eol,ALICE:2019hdt,ALICE:2020mfd,ALICE:2021cpv, ALICE:2021cyj,ALICE:2022wpn,LKana}.

The study of hadronic interactions involving charm mesons (D, \Dstar) has gained significant interest after the observation of the charm-strange meson \Dsexc{0}{2317}~\cite{BaBar:2003oey,Belle:2003guh,CLEO:2003ggt}, whose mass lies significantly below the quark
model~\cite{Godfrey:1985xj} predictions ($m_\mathrm{experiment} - m_\mathrm{quark~model} \approx 100~\MeVc^2$), preventing its accommodation in simple constituent quark models~\cite{Ortega:2016mms}. The puzzle of the \Dsexc{0}{2317} low mass has led to a range of theories, such as those based on the concepts of conventional charm-strange mesons with coupled-channel impacts~\cite{vanBeveren:2003kd, Godfrey:2003kg, Bardeen:2003kt, Fayyazuddin:2003aa, Colangelo:2003vg, Colangelo:2005hv, Lu:2006ry}, or of \DorDstarNoSign{K} molecule~\cite{Barnes:2003dj,Szczepaniak:2003vy,Chen:2004dy,Guo:2006fu,Guo:2006rp}, or of a tetraquark state composed of $\mathrm{cq\overline{s}\overline{q}}$ (anti)quarks~\cite{Cheng:2003kg,Bracco:2005kt,Dmitrasinovic:2005gc}. Models based on a mixture of tetraquark and molecular states were also proposed~\cite{Browder:2003fk, Vijande:2006hj}. In recent years, several exotic hadrons with charm-quark content have been discovered, such as the \Chic~\cite{Belle:2003nnu}, \Tcc~\cite{LHCb:2021vvq,LHCb:2021auc}, \Pc{4312}, \Pc{4440}, and \Pc{4457}~\cite{LHCb:2015yax,LHCb:2019kea,Godfrey:2003kg} states. Similarly to the \Dsexc{0}{2317}, these states can be interpreted as \DDstarbar, \DDstar, or \SigmacDbar, \SigmacDstarbar molecular states, or compact multiquark states~\cite{Chen:2016qju,Brambilla:2019esw,Kamiya:2022thy,Yalikun:2021bfm}.
The observation of potential molecular states is, however, not the only measurement that challenges the charm-hadron spectrum in terms of the conventional quark model. In fact, the masses of the non-strange \Dexc{0}{2300} and \DexcNostar{1}{2430} charm mesons~\cite{Belle:2003nsh,LHCb:2015klp,LHCb:2016lxy} are very similar to the corresponding states in the charm-strange spectrum, \Dsexc{0}{2317} and \DsexcNostar{1}{2460}~\cite{BaBar:2003oey,CLEO:2003ggt,Godfrey:2003kg}, while they are expected to be smaller. When combining chiral effective field theory with quantum chromodynamics calculations on the lattice, all low-energy open heavy-flavor mesonic states with positive parity can be classified as hadronic molecules. In this framework, pions, kaons, and $\upeta$ mesons arise as Goldstone bosons and, by computing the \DPi, \Deta, and \DsAntiK coupled-channel scatterings, a bound state with a large coupling to the \DPi channel is obtained at a mass that corresponds to the \Dexc{0}{2300} state~\cite{Du:2017zvv,Albaladejo:2016lbb,Du:2020pui,Asokan:2022usm,Albaladejo:2023pzq}. Nevertheless, their structures remain uncertain owing to the lack of direct experimental information on the residual strong interaction between charm and light hadrons.
These measurements are particularly challenging because conventional scattering experiments with charm hadrons are restricted by their short lifetime.
Only recently, the residual strong final state interaction involving charm hadrons became experimentally accessible thanks to the femtoscopy technique.
The first study of the strong interaction between charm mesons and nucleons (\pD) was published by the ALICE Collaboration in Ref.~\cite{ALICE:2022enj}, proving the feasibility of applying the femtoscopy technique to the charm sector.

The knowledge of interactions between charm particles and light-flavor hadrons is also essential for the study of ultrarelativistic heavy-ion collisions. In these collisions, a color-deconfined state of matter, called quark--gluon plasma (QGP), is formed~\cite{Braun-Munzinger:2015hba, STAR:2005gfr, PHOBOS:2004zne, PHENIX:2004vcz, BRAHMS:2004adc}. Due to the early production, charm quarks are recognized as ideal probes of the QGP and measurements of the yields and angular anisotropies of charm hadrons can be used to infer information about the QGP properties~\cite{ALICE:2020iug,ALICE:2021rxa}.
However, during the hadronic phase following the deconfined state of the system, the charm hadrons can interact with the other particles produced in the collision, which are mainly light-flavor hadrons, via elastic and inelastic processes. These interactions modify the momentum and angular distributions of heavy-flavor hadrons in heavy-ion collisions. Therefore, the scattering parameters of the charm hadrons with light-flavor hadrons, in particular pions and kaons, must be determined to disentangle this effect from those related to the QGP formation~\cite{He:2011yi}.

In this article, the first measurement of the residual strong interaction between non-strange charm and light-flavor mesons via the femtoscopy technique is presented. This method relies on the fact that particles with similar momentum, hence small relative momentum, can interact with each other strongly, if they are emitted at small relative distance. The momentum correlation functions of the charm mesons \Dplus and \Dstarp with charged pions and kaons, also simply referred to as light-flavor mesons in the following, are measured for all charge combinations in pp collisions at \thirteen. 
Section~\ref{sec:analysis} contains the description of the experimental apparatus, the selection of charm and light-flavor mesons, as well as the single-particle properties (e.g. purity), which are later needed to extract the final results from the raw experimental data. The measurement of the correlation functions is described in Section~\ref{sec:cf}, while the evaluation of the systematic uncertainties is discussed in Section~\ref{sec:syst}. Finally, the results are presented and compared to model calculations in Section~\ref{sec:results}.

\section{Event and particle selection}
\label{sec:analysis}

This analysis is performed on a data sample of pp collisions at \thirteen collected with the ALICE~\cite{ALICE:2022wpn} experiment during the LHC Run~2 data-taking period.
The events are selected employing a high-multiplicity (HM) trigger. The multiplicity is estimated using the V0 detector, which consists of an array of scintillators located at forward (${2.8 < \eta < 5.1}$) and backward (${-3.7 < \eta < -1.7}$) pseudorapidity~\cite{VZERO}.
The multiplicity estimator is the V0 amplitude, which is related to the energy deposited by ionizing particles in the V0 detector. The triggered events correspond to the 0--0.17\% percentile of the inelastic events with the highest V0 amplitude and with at least one charged track in the range $|\eta| < 1$ (${\text{INEL} > 0}$).
The resulting HM dataset consists of approximately $1.0 \times 10^9$ inelastic \pp collisions with, on average, 30 charged particles per event in the pseudorapidity interval \etarange{0.5}~\cite{ALICE:2020mfd}.
Charged-particle tracks are reconstructed using both the Inner Tracking System (ITS)~\cite{ALICEITS} and the Time Projection Chamber (TPC)~\cite{ALICETPC}, which are embedded in a uniform magnetic field of \SI{0.5}{\tesla} along the beam direction. They cover the full azimuthal angle and the pseudorapidity interval \etarange{0.9}. The position of the primary vertex is obtained from the reconstructed tracks, and the particle identification (PID) is performed employing both the TPC and the Time-of-Flight (TOF)~\cite{ALICETOF} detectors.

The \pythia 8.243 event generator~\cite{PYTHIA} is used in the Monte Carlo (MC) simulations.
The generated particles are transported through a simulation of the ALICE apparatus using GEANT 3~\cite{GEANT3}. Events and tracks are reconstructed employing the same algorithms as used for real collision data~\cite{ALICE} and a selection on large charged-particle multiplicities is applied to mimic the effect of the HM trigger.

\subsection{Light-meson selection}\label{sec:lightselections}

The K$^+$ and $\uppi^+$ candidates are identified using PID information provided by the TPC and TOF, via the specific energy loss $\mathrm{d}E/\mathrm{d}x$ and time-of-flight, respectively. For each track, the deviation of the measured quantity with respect to the expected value for a particular particle-species hypothesis in terms of units of the detector resolution is computed and denoted as ${n_\sigma^\mathrm{TPC/TOF}}$. 
Pion candidates with transverse momentum $\pt < 0.5~\GeVc$ are identified using only the TPC $\mathrm{d}E/\mathrm{d}x$ signal via a selection of $|n^\TPC_\sigma(\uppi)|<3$. For larger $\pT$ the PID information of TPC and TOF is combined into ${n_\sigma^\mathrm{comb} = \sqrt{(n^\TPC_\sigma)^2 + (n^\TOF_\sigma) ^2}}$ and a selection of $n_\sigma^\mathrm{comb} <3$ is applied. Tracks with $\pt > 0.5~\GeVc$ which do not have a TOF signal are discarded. The PID selection of the kaon candidates is performed similarly with an additional more complex set of selections on $n^\TPC_\sigma$ and $n^\mathrm{comb}_\sigma$, not only for kaons but also for electrons and pions, in order to suppress possible contamination to the kaon sample in specific momentum regions~\cite{LKana}.

The pion and kaon candidates are selected in the \pt ranges [0.14, 4.0] and [0.15, 2.15]~$\GeVc$, respectively. The lower limit is imposed to suppress the light-meson candidates stemming from interactions with the detector material.
The tracks are required to be reconstructed from more than 80 clusters in the TPC to assure a good quality of the track, good \pt resolution at large momenta, as well as to remove fake tracks from the sample.
In addition, the candidates are selected within a pseudorapidity range of $|\eta|< 0.8$.
To suppress the contribution of particles coming from weak decays or interactions with the detector material, a selection on the distance of closest approach (DCA) to the primary vertex in the transverse plane $xy$ and along the beam axis direction $z$ is applied. For kaons, $\mathrm{DCA}_{xy}^{\mathrm{K}} < 0.1~\centm$ and $\mathrm{DCA}_{z}^{\mathrm{K}} < 0.2~\centm$ are required, while for pions $\mathrm{DCA}_{xy,z}^\uppi < 0.3~\centm$.

The purity of the pion and kaon samples, defined as the ratio of the correctly identified particles over the total number of candidates, is computed as a function of \pt using MC simulations, and is reweighted by the \pt distribution of the pion or kaon candidates that form a pair with \DorDstarp mesons at low relative momentum. It is found to be 99\% for pions and 98\% for kaons. 

The particles can be classified according to their origin: the ones that do not come from interactions with the material of the detector are classified as primary or secondary, according to the ALICE definition~\cite{ALICE:2017hcy}. The fraction of each contribution is estimated with a template fit to the DCA distribution. The templates for the DCA distributions of primary particles, secondaries from weak decays, and secondaries from interactions in the material are obtained from MC simulations. The primary fractions are found to be 99.5\% and 99.8\% for pions and kaons, respectively. A portion of identified primary light-flavor mesons comes, however, from the strong decay of long-lived resonances ($c\tau > 5~\mathrm{fm}$). As the fractions of this contribution cannot be determined via DCA template fits, they are estimated with the \fist statistical hadronization model~\cite{Vovchenko:2019pjl}. The resonances that contribute the most to the pion yield are the $\upeta$ and $\upomega$ mesons, while in the case of kaons it is the $\upphi$ meson. The resulting primary fractions of pions and kaons, subtracted of the contribution of such long-lived resonances, are found to be about 88\% and 94\%, respectively. These values are used in the following analysis as primary fractions.

\subsection{Charm-meson selection}
\label{sec:analysisCharm}
The \Dplus, \Dstarp, and \Dzero candidates are reconstructed via the hadronic decay channels \DplustoKpipi, \DstartoDpi, followed by \DzerotoKpi, and their charge conjugates. The branching ratios (BR) of the considered \Dplus, \Dstarp, and \Dzero decays are $\mathrm{BR}=(9.38\pm0.16)\%$, $\mathrm{BR}=(67.7\pm0.5)\%$, and $\mathrm{BR} = (3.947\pm0.030)\%$, respectively~\cite{ParticleDataGroup:2022pth}. The tracks fulfilling a set of standard quality selections~\cite{ALICE:2022enj} are combined with the correct charge signs to build \Dplus- and \Dstarp-meson candidates. The obtained sample of charm-meson candidates consists of three different classes: candidates that result from the combination of uncorrelated pions and kaons form the \textit{combinatorial background}, charm mesons that come from the hadronization of a charm quark or the decay of excited open-charm or charmonium states, which are referred to as \textit{prompt}, and \DorDstarp mesons that come from the decay of beauty hadrons, which are referred to as \textit{non-prompt}.

To separate the prompt, non-prompt, and combinatorial background contributions, the decay-vertex topology, in combination with the PID information is used. The mean proper decay length of \Dpm and \Dzero mesons is about 312~\mum and 123~\mum, respectively, while for beauty hadrons it is close to 500~\mum~\cite{ParticleDataGroup:2022pth}.
Topological variables, such as the DCA of the charm meson candidate, the \Dplus (\Dzero) decay length, and the cosine of the pointing angle, namely the angle between the \Dplus (\Dzero) momentum and the line that passes through the primary and secondary vertices, are exploited by a multi-class machine learning (ML) algorithm based on Boosted Decision Trees (BDT). The ML model, provided by the XGBoost library~\cite{Chen:2016:XST:2939672.2939785,barioglio_luca_2022_7014886}, is trained using labeled examples of candidates of each class. The samples of prompt and non-prompt \Dplus and \Dstarp mesons are obtained from a \pythia 8 simulation with enhanced production of heavy-flavor hadrons, where only events that contain a $\mathrm{c\overline{c}}$ or $\mathrm{b\overline{b}}$ pair are selected, and the charm mesons are forced to decay in the hadronic decay channels of interest for the analysis. The background sample for \Dplus is obtained from the data by selecting the sidebands of the candidate invariant-mass distribution. For \Dstarp mesons, the right sideband of the invariant-mass difference $\Delta M = M(\mathrm{K}\uppi\uppi) - M(\mathrm{K}\uppi)$ is used. To prepare the sample for the training, loose selections on the PID and decay-vertex topology are applied. The training is performed in several \pt intervals.
Then, the model is applied to the data, assigning scores to each candidate, which are related to the probabilities that the candidate belongs to each of the three classes.
To suppress the combinatorial background and enhance the prompt contribution in the sample, candidates with a low background-score and high prompt-score are selected; the selections are chosen such that they maximize the expected significance and purity.

\begin{figure}[!tb]
    \centering
    \includegraphics[width=0.49\linewidth]{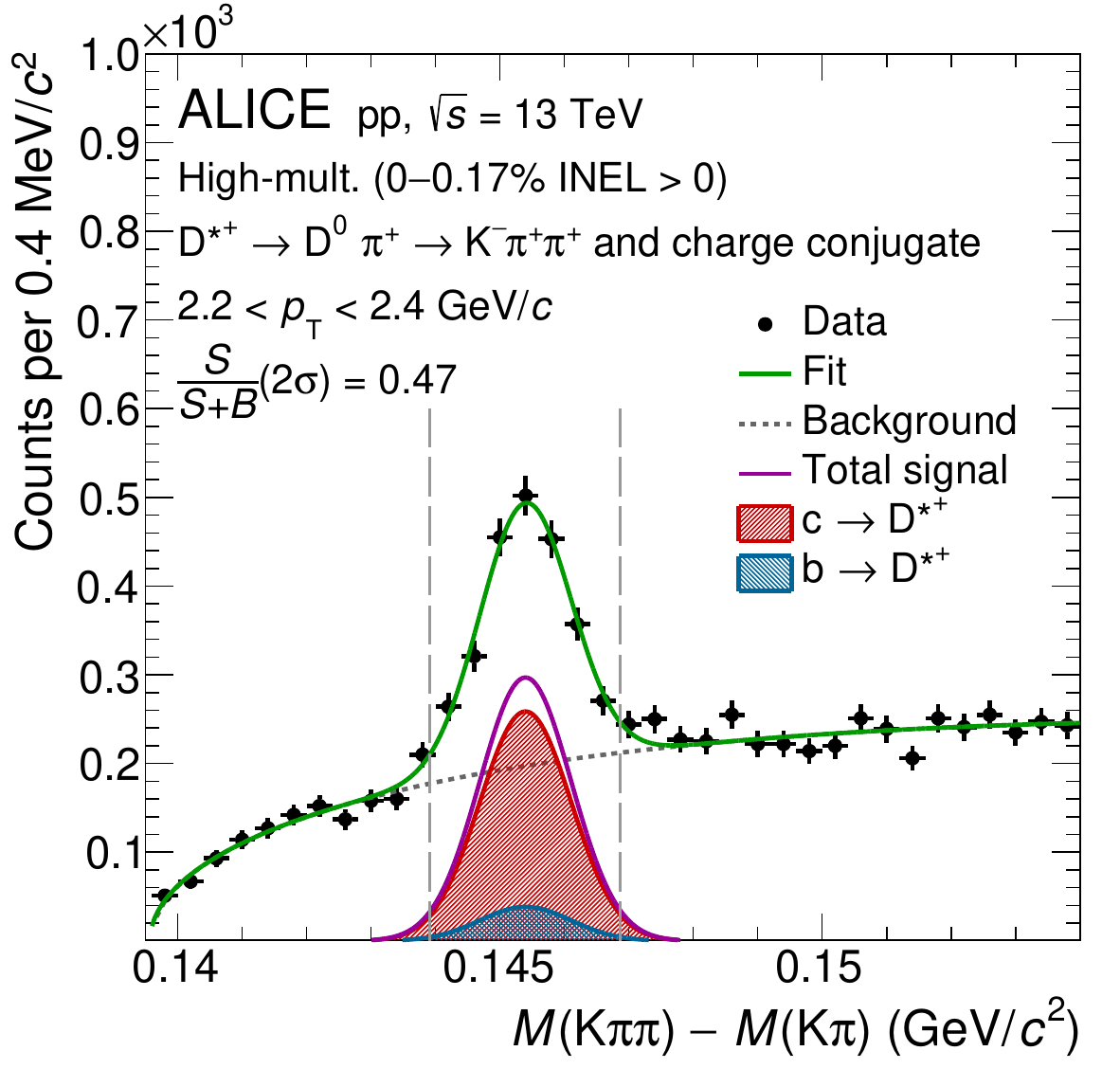}
    \includegraphics[width=0.49\linewidth]{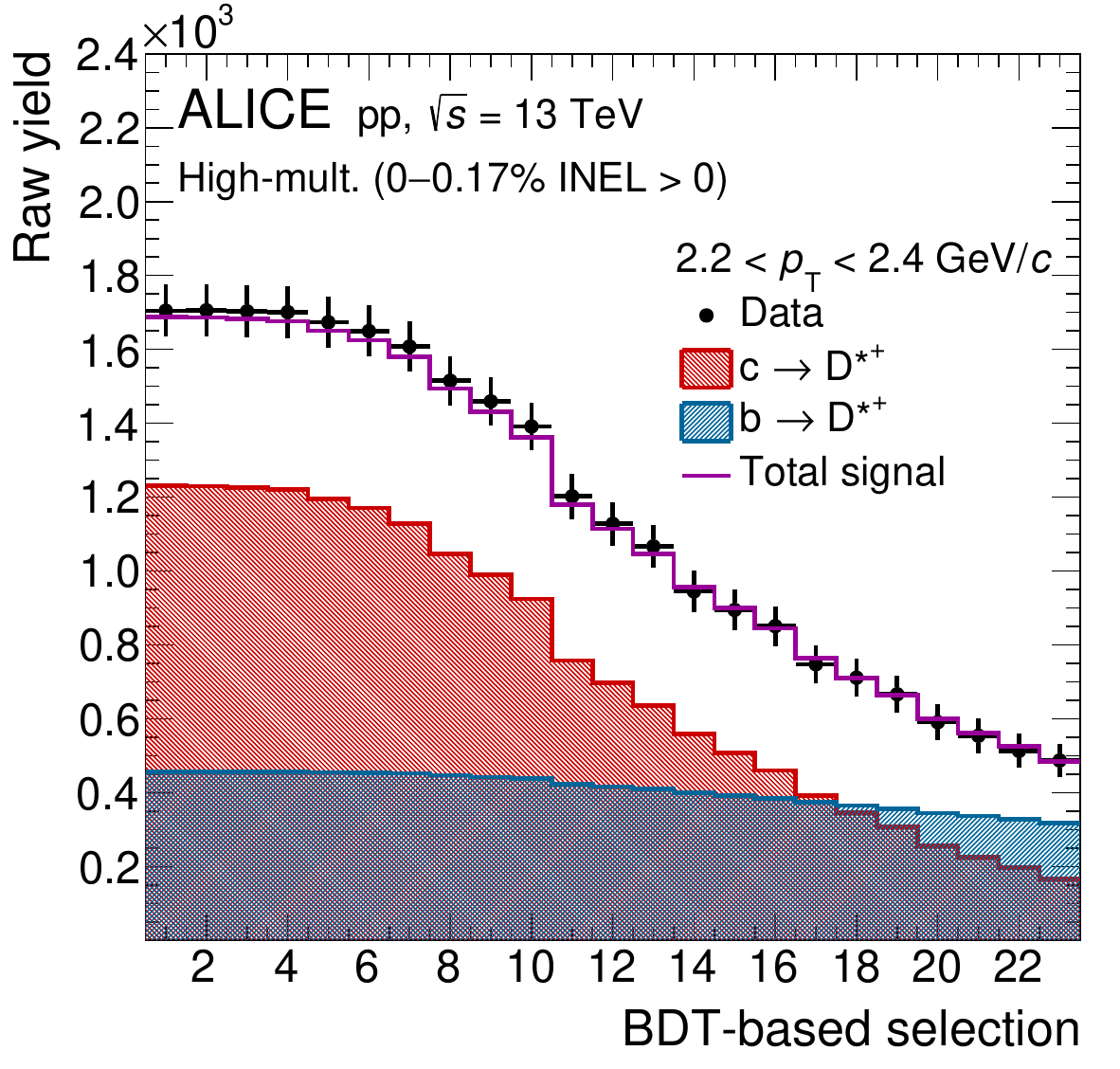}
    \caption{Left: distribution of invariant-mass difference for \Dstarp candidates in the $2.2<\pt<2.4~\GeVc$ interval. The green solid line shows the total fit function and the gray dotted line the combinatorial background. The contributions of \Dstarp mesons originating from charm hadronization and beauty-hadron decays are obtained with the method relying on the definition of different selection criteria, as explained in the text. Right: example of raw-yield distribution as a function of the BDT-based selection for the $2.2<\pt<2.4~\GeVc$ interval, employed in the procedure adopted for the determination of the fraction of \Dstarp originating from beauty-hadron decays.}
    \label{fig:massDstar}
\end{figure}

The fraction of non-prompt candidates present in the sample is estimated with a data-driven procedure that relies on the fact that the prompt selection efficiencies change differently to the non-prompt ones when the selection on the ML scores is changed. For each selection $i$ on the ML scores, the raw yield $Y_\mathrm{i}$ of charm-meson candidates is extracted via a fit to the invariant-mass distribution of the charm-meson candidates. The fit function is the sum of a Gaussian, for the description of the signal, and an exponential or an exponential multiplied by a power law for the description of the background in the case of \Dplus and \Dstarp mesons, respectively. The left panel of Fig.~\ref{fig:massDstar} shows an example of fit to the $\Delta M$ distribution of \Dstarp candidates with $2.2<\pt<2.4~\GeVc$.
The raw yield is related to the corrected yields of prompt ($N_\mathrm{prompt}$) and non-prompt ($N_\mathrm{non\text{-}prompt}$) mesons via
\begin{equation}
    \delta_\mathrm{i} = Y_\mathrm{i} - (\mathrm{Acc}\times \epsilon)_{\mathrm{prompt}, i}\times N_\mathrm{prompt} - (\mathrm{Acc}\times \epsilon)_{\text{non-prompt}, i}\times N_\text{non-prompt},
\end{equation}
where $(\mathrm{Acc} \times \epsilon)_\text{prompt/non-prompt}$ is the product of acceptance and efficiency for each selection and $\delta_\mathrm{i}$ are the residuals that account for the equation not holding exactly because of the uncertainties. The definition of multiple sets of selections leads to an overdetermined system of equations, out of which the corrected yields can be extracted via a $\chi^2$ minimization. Further details are provided in Ref.~\cite{ALICE:2021mgk}. An example of a raw-yield distribution as a function of the BDT-based selection used in the minimization procedure for \Dstarp mesons with $2.2<\pt<2.4~\GeVc$ is shown in the right panel of Fig.~\ref{fig:massDstar}. The leftmost data point of the distribution represents the raw yield corresponding to the loosest selection on the BDT output related to the candidate probability of being a non-prompt \Dstarp meson, while the rightmost one corresponds to the strictest selection, which is expected to preferentially select non-prompt \Dstarp mesons. The prompt and non-prompt components obtained from the minimization procedure are represented by the red and blue filled histograms, respectively.
The non-prompt fraction extracted in \pT intervals is reweighted with the \pt distribution of the \DorDstarp mesons that form pairs at low \kstar. The extracted non-prompt fractions are $(7.2 \pm 0.2)\%$ for \DPi and \DK, and $(7.7\pm 1.3)\%$ for \DstarPi and \DstarK.

The prompt component of the \Dplus-meson sample also includes mesons that come from the decay of excited charm states. The main contribution comes from the decay of the \Dstarp mesons, via the $\Dstarpm \to \Dpm + \uppi^0$ and $\Dstarpm \to \Dpm + \gamma$ decays, that have a branching ratio of $(30.7\pm0.5)\%$ and $(1.6\pm0.4)\%$, respectively~\cite{ParticleDataGroup:2022pth}. Since the strong final-state interaction (FSI) is only accessible via the study of the primary particles, the \Dplus mesons that result from the decay of charm resonances represent a source of background. Unlike the contribution of \Dplus mesons from beauty-hadron decays, it is not possible to experimentally separate it with the procedure described above, due to the short lifetime of the \Dstarp resonances ($c\tau \approx 2400~\fm$)~\cite{ParticleDataGroup:2022pth}. The fraction of \Dplus mesons originating from \Dstarp decays is estimated in Ref.~\cite{ALICE:2022enj}, employing the production cross sections of \Dplus and \Dstarp mesons in pp collisions at \five~\cite{ALICE:2019nxm,ALICE:2021mgk} and a simulation with \pythia~8.2 for the description of the $\Dstarpm \to \Dpm + \mathrm{X}$ decay kinematics. It is estimated to be $(27.6\pm1.3~(\mathrm{stat.})\pm 2.4~(\mathrm{syst.}))$\%.

To obtain a high-purity sample of \DorDstarp-meson candidates, the following procedure is used.
The distribution of the invariant mass of the \Dplus-meson candidates and invariant-mass difference of the \Dstarp-meson candidates is fitted in several \pt intervals, from 1 to $10~\GeVc$. The sample of \DorDstarp mesons used for the analysis is obtained by applying a selection to the invariant mass of the candidates, which is defined by a 2$\sigma$ window around the nominal mass, $M_{\Dpm}=1869.66\pm0.05~\MeVc^2$ and $M_{\Dstarpm}=2010.26\pm0.05~\MeVc^2$~\cite{ParticleDataGroup:2022pth}, where $\sigma$ is the width of the fitted Gaussian. This selection range is represented by the vertical dashed lines in Fig.~\ref{fig:massDstar}. The purity is computed as the ratio of the signal candidates over the total number of candidates in this invariant-mass range, where the number of signal candidates is extracted with a fit to the invariant-mass distribution. 
This results in a \pt-integrated purity of around 71\% for \Dplus mesons and 67\% for \Dstarp mesons.

\section{The correlation function}
\label{sec:cf}

In this analysis, the interaction between the charm mesons \DorDstarNoSign and the light-flavor mesons $\uppi$ and K is investigated employing the correlation function~\cite{Lisa:2005dd}, defined as
\begin{equation}\label{eq:cf}
C(\kstar)= \mathcal{N}\times\frac{N_{\rm{same}}(\kstar)}{N_{\rm{mixed}}(\kstar)}, 
\end{equation}
where $\kstar = \frac{1}{2}\times|\mathbf{p}^*_1-\mathbf{p}^*_2|$ is the relative momentum of two particles with momentum $\pmb{p}_1$ and $\pmb{p}_2$ in the pair rest frame, denoted by the asterisk, $\mathcal{N}$ is a normalization constant, and $N_\text{same (mixed)}(\kstar)$ is the \kstar distribution of the pairs measured in the same (mixed) events. The mixed-event distribution, which does not contain any effect of the strong FSI, reflects the phase space of the underlying event. Therefore, it serves as a reference to which the same-event distribution can be compared in order to extract information on the strong FSI of a specific system. 
To ensure a good quality of the reference sample, $N_\mathrm{mixed}$, the mixing is performed only between events with similar multiplicity and primary-vertex position~\cite{ALICE:2019buq,ALICE:2019gcn,ALICE:2020mfd}. 
As the same (mixed) event distributions of the pairs are found to be compatible with the ones of the respective charge conjugates, they are combined in order to enhance the statistical precision. In the following, same-charge $\mathrm{D}^{(*)}\mathrm{X}$ refers to $\mathrm{D}^{(*)+}\mathrm{X}^{+}\oplus \mathrm{D}^{(*)-}\mathrm{X}^{-}$ pairs, while opposite-charge $\mathrm{D}^{(*)}\mathrm{X}$ refers to $\mathrm{D}^{(*)+}\mathrm{X}^{-}\oplus \mathrm{D}^{(*)-}\mathrm{X}^{+}$ pairs, where X is either K or $\uppi$.
The normalization constant $\mathcal{N}$ is chosen such that the mean value of the correlation function equals unity in a given range at large \kstar, where the particles are not close enough in momentum space to experience FSI. 
The number of pairs and the normalization range for the different channels are reported in Table~\ref{tab:pairs}. The latter are chosen according to the shape of the same (mixed) event distributions, which decreases and flattens out at different \kstar regions depending on the involved light-flavor meson. 
The experimental correlation functions are computed in \kstar intervals of $50~\MeVc$ and the horizontal position of each data point is the average of the \kstar distribution of the mixed event in the corresponding \kstar interval. The effect of the finite momentum resolution of the ALICE detector on data is found to be negligible.

\begin{table}
\begin{center} 
\caption{Number of pairs with small relative momenta, where final-state effects become relevant, and in the full \kstar range, as well as the normalization range for the individual particle pair combinations under investigation.}

\renewcommand*{\arraystretch}{1.2}

\begin{tabular}{c | c | c | c}
\multirow{2}{*}{{ Pair}} & \multicolumn{2}{c|}{Number of pairs in $N_\mathrm{same}(\kstar)$} & \multirow{2}{*}{Normalization range}\\ \cline{2-3}
 &Total & $\kstar<200~\MeVc$ \\ 
 \hline
\hline  
\DPisc& $\num{3.0e6}$ &$\num{2.0e5}$& \multirow{2}{*}{$\kstar\in
[1.0,~1.5]$\,$\GeVc$} \\ 
\DPioc &$\num{2.9e6}$ & $\num{2.1e5}$\\ 
\hline
\DKsc& $\num{1.7e5}$ &\num{1.9e3} & \multirow{2}{*}{$\kstar\in[1.5,~2.0]$\,$\GeVc$}\\ 
\DKoc & $\num{1.6e5}$ &\num{2.2e3}&\\  \hline
\DstarPisc & $4.7 \times 10^5$ & $3.3\times 10^4$ & \multirow{2}{*}{$\kstar\in[1.5,~2.0]$\,$\GeVc$}\\
\DstarPioc & $4.8 \times 10^5$ & $3.4\times 10^4$\\
\hline
\DstarKsc & $4.9\times 10^4$ & 479 & \multirow{2}{*}{$\kstar\in[1.5,~2.0]$\,$\GeVc$}\\  
\DstarKoc & $4.8\times 10^4$ & 477\\
\hline
\hline

\end{tabular}
\label{tab:pairs}
\end{center}
\end{table}

The experimental correlation functions involving \Dplus and light-flavor mesons, obtained from Eq.~\ref{eq:cf} are shown in the left panels of Figs.~\ref{fig:bkg_pions} and~\ref{fig:bkg_kaons}. They are raw quantities, which can be decomposed as
\begin{align}\label{eq:rawcf}
    C_\mathrm{raw}(\kstar) = C_\mathrm{femto}(\kstar) \times C_\text{non-femto} (\kstar),
\end{align}
where $C_{\rm{femto}}(\kstar) = \sum_{\mathrm{i},\mathrm{j}} \lambda_{\mathrm{i},\mathrm{j}} \times C_{\mathrm{i},\mathrm{j}}(\kstar)$, with $C_{\mathrm{i},\mathrm{j}}(\kstar)$ arising from the FSI between the $i$-th and $j$-th components of the two particle species involved in the analysis, namely primary, secondary, and misidentified particles. Each of these contributions is weighted according to so-called $\lambda$ parameters, which are computed as $\lambda_{\mathrm{ij}} = p_\mathrm{i}p_\mathrm{j}f_\mathrm{i}f_\mathrm{j}$ where $p_{\mathrm{i},\mathrm{j}}$ and $f_{\mathrm{i}, \mathrm{j}}$ are, respectively, the purities and primary (secondary) fractions of the $i$-th and $j$-th contributions to the particle samples, discussed in Section~\ref{sec:lightselections}.
The contribution to $C_{\rm{femto}}(\kstar)$, that only includes primary signal particles, is also referred to as genuine correlation function $C_{\rm{gen}}(\kstar)$ and is used to extract the relevant physics information about the strong FSI for the pair of interest. A detailed discussion on the different contributions to $C_\text{femto} (\kstar)$ can be found in Section~\ref{subsec:FSI}. 
The remaining residual backgrounds, not related to FSI, are included in the term $C_\text{non-femto} (\kstar)$, which is discussed in Section~\ref{subsec:residual}.

\subsection{Contributions related to FSI}
\label{subsec:FSI}

There are several contributions to $C_\text{femto} (\kstar)$ in Eq.~\ref{eq:rawcf} in the case of \DorDstarp and light-flavor mesons.
When it is not possible to constrain them experimentally, these contributions can be modeled using the Koonin--Pratt equation~\cite{Lisa:2005dd}
\begin{equation}
C(\kstar)= \int \, \mathrm{d}^3 r^* S(\pmb{r}^*) |\psi(\pmb{r}^*,\pmb{k}^*)|^2,
\label{eq:CFsourcewf}
\end{equation}
where the so-called source function $S(\pmb{r}^*)$ contains the distribution of the relative distance in the pair rest frame, and $\psi(\pmb{r}^*,\pmb{k}^*)$ denotes the two-particle wave function, which contains the interaction. Together they determine the shape of the correlation function, which is sensitive to the strong FSI at small $\kstar<200~\MeVc$, also denoted as femtoscopic region.

The source is constrained from the core--resonance model~\cite{ALICE:2020ibs}, which is based on the hypothesis of a common emission source of all hadrons~\cite{KorwieserSource} and is anchored to p--p correlation data in pp collisions. The model is characterized by a \mt-dependent Gaussian core of width $r_{\rm{core}}$, from which all primordial particles, which are created directly during the hadronization process, and do not stem from an intermediate decay, are emitted.  Therefore, by measuring the \mt of the reconstructed particle pairs with small \kstar it is possible to obtain the respective core radius from a parameterization of the p--p data used in the model, following several previous femtoscopic analyses~\cite{ALICE:2022enj, ALICE:2021cyj, ALICE:2021cpv,ALICE:2021njx,ALICE:2020mfd, ALICE:2019buq, ALICE:2022yyh}.
The mean \mt of \DorDstarNoSignPi pairs with $\kstar<200~\MeVc$ is about $2.55~\GeVcc$, while it is approximately $2.66~\GeVcc$ for \DorDstarNoSignK pairs. This leads to core radii of $r^{\DorDstarNoSignPi}_{\rm{core}}=0.82^{+0.07}_{-0.07}~$ fm and $r^{\DorDstarNoSignK}_{\rm{core}}=0.81^{+0.08}_{-0.07}~$ fm for \DorDstarNoSignPi and \DorDstarNoSignK pairs respectively.
However, also short-lived resonances feeding into the yields of the particles of interest have to be considered, as they lead to an effective enlargement of the source. This is accounted for in the core--resonance model by fixing the yields of the resonances and employing an event generator to model their propagation and relative spatial orientation. At large \rstar these resonances lead to an exponential tail in the Gaussian-shaped source distributions obtained from the model for both \DorDstarNoSignK and \DorDstarNoSignPi. Therefore, the effective source employed in this analysis is obtained by parameterizing the distributions with two Gaussian sources of width $r_{\rm{eff}}^i$, which are combined with the weight $w$, leading to $S_\mathrm{eff}(\rstar)=w S_1(\rstar)+(1-w)S_2(\rstar)$. The values of the source parameters can be found in Table~\ref{tab:effrad}. Employing $S_{\rm{eff}}(\rstar)$ as source function in Eq.~\ref{eq:CFsourcewf} ultimately leads to two properly weighted correlation functions with the respective Gaussian sources, $S_1(\rstar)$ and $S_2(\rstar)$.\\
The two-particle wave function $\psi(\pmb{r}^*,\pmb{k}^*)$ can be obtained by numerically solving the Schrödinger equation for a given interaction potential, for example by employing CATS~\cite{Mihaylov:2018rva}, a correlation analysis tool using the Schrödinger equation. 
\begin{table}[!b]
\centering
\caption{Parameters of the effective source $S_\mathrm{eff}(\rstar)$, which is given by the weighted sum of two Gaussian distributions of width $r^i_{\text{eff}}$ and used in the modeling of the correlation functions. The difference between the \DorDstarNoSignPi and \DorDstarNoSignK systems is due to the different transverse mass of the systems as well as resonances feeding into the light-flavor mesons.}
\renewcommand*{\arraystretch}{1.2}
\begin{tabular}{c | c |c |c }
Pair & $w$ & $r^1_{\mathrm{eff}}$ [fm] & $r^2_{\mathrm{eff}}$ [fm] \\
\hline
\hline
\DorDstarNoSignK & $0.78^{+0.02}_{-0.01}$ &$0.86^{+0.09}_{-0.07}$& $2.03^{+0.19}_{-0.12}$\\
\DorDstarNoSignPi & $0.66^{+0.03}_{-0.02}$ &$0.97^{+0.09}_{-0.08}$& $2.52^{+0.36}_{-0.20}$\\
\hline
\hline
\end{tabular} 
\label{tab:effrad}
\end{table}

\begin{table}[]
    \centering
    \caption{List of $\lambda$ parameters, which quantify the individual contributions to the different raw correlation functions investigated in this paper.}
    \renewcommand*{\arraystretch}{1.2}
    \begin{tabular}{c|c|c|c|c|c}
         & D$\uppi$ & DK   & D$^*\uppi$ & \DstarKsc & \DstarKoc \\
         \hline\hline                                                                             
         $\lambda_\mathrm{gen}$        & 0.40     & 0.43 & 0.80       & 0.55        & 0.59      \\
         $\lambda_\mathrm{SB}$         & 0.29     & 0.29 & n.a.       & 0.32        & 0.28      \\
         $\lambda_\mathrm{flat}$       & 0.14     & 0.10 & 0.20       & 0.13        & 0.13      \\
         $\lambda_{\mathrm{D_{\mathrm{D} \leftarrow \Dstar}}}$  & 0.17     & 0.18 & n.a.       & n.a.        & n.a.      \\
         \hline\hline

    \end{tabular}
    \label{tab:lambdapars}
\end{table}
The relevant contribution to $C_{\rm{femto}}(\kstar)$, needed to extract information of the strong FSI between \DorDstarNoSignPi and \DorDstarNoSignK, is the genuine correlation function $C_{\rm{gen}}(\kstar)$, which is associated to primary light-flavor mesons and signal \DorDstarp candidates.

As the \DorDstarp-meson samples are not pure, the correlation between combinatorial background candidates and light-flavor mesons has to be taken into account, which arises from the interaction between the light-flavor mesons and the particles from which the background \DorDstarp-meson candidate is built from~\cite{DelGrande:2021mju}.
It is estimated using a data-driven approach, where pions or kaons are paired with a pure sample of background \DorDstarp mesons, obtained from the sidebands of the invariant-mass intervals outside the \DorDstarp-meson signal region. The resulting correlation function is referred to as $C_{\mathrm{SB}}(\kstar)$

For the \Dplus mesons, the sideband intervals start at 5~$\sigma_\mathrm{D}$ away from the nominal mass and extend for $200~\MeVcc$. The $\sigma_\mathrm{D}$ corresponds to the width of the Gaussian function describing the signal peak and is determined via a fit to the invariant-mass distribution, considering its \pt dependence. For the \Dstarp mesons, the selection is analogous except that, instead of the invariant mass, the invariant-mass difference $M(\mathrm{K}\uppi\uppi) - M(\mathrm{K}\uppi)$ is used, and only the right sideband is considered.

Since a contamination from \Dstarp-meson is expected in the \Dplus-meson sideband sample, due to $\mathrm{D^{*+} \to D^0 \uppi^+}$ and subsequent $\mathrm{D^0\to K^-\uppi^+}$ decays, the invariant-mass interval $[1.992,~2.028]~\MeVc^2$ is excluded. This corresponds to $2.5~\sigma_{\Dstar}$ around the \Dstarp mass. The correlation functions obtained from the left and right sidebands are compatible within the uncertainties and combined as a weighted average, considering the relative abundances of background in the left and right half of the \Dplus-meson signal region.
The correction of the combinatorial \DstarPi correlation function requires a different approach with respect to the traditional
sideband method. This is due to the presence of an additional source of correlated background that arises from the correlation of a soft pion of a real \Dstarp decay with a background \Dstarp candidate formed by the \Dzero meson coming from the same \Dstarp decay of the soft pion and an unrelated pion.
Such a correlation results in a peak in the correlation function at $\kstar\approx 40~\MeVc$, which cannot be removed via pair- or particle-level selections since the particle’s origin is not known in data. For this reason, the correction for the combinatorial background cannot be carried out via a sideband analysis. Instead, the background-corrected correlation function is directly computed as
\begin{equation}
    C_\mathrm{raw}^\prime(\kstar) = \mathcal{N} \frac{p_\mathrm{same}(\kstar) N_\mathrm{same}(\kstar)}{p_\mathrm{mixed}(\kstar) N_\mathrm{mixed}(\kstar)},
    \label{eq:sblessapproach}
\end{equation}
where $p_\mathrm{same/mixed}(\kstar)$ is the purity of the \Dstarp-meson sample, calculated in the same- and mixed-events, as a function of \kstar.
Since the peak in the correlation function comes from the combinatorial background of the \Dstarp-meson candidates, a reweighting by the purity removes by construction the artifact at $\kstar\approx 40~\MeVc$.
The opposite-charge \DstarK correlation function is affected by a similar issue since the \Dzero meson decays into $\mathrm{K}^-$ via \DzerotoKpi. However, in this case, the peak associated with the correlated background is found to be at $\kstar \approx 600~\MeVc$, outside the femtoscopic region. As the correlation function above $200~\MeVc$ does not carry information about the strong FSI, the traditional sideband method is used to correct for the combinatorial background.

As already discussed in Section~\ref{sec:analysisCharm}, a significant fraction of the \Dplus mesons is produced from the decays of charm-hadron resonances. As this contribution cannot be separated experimentally, it is modeled using the Koonin–Pratt formalism with Coulomb potential, which is found to adequately describe the experimental correlation functions involving \Dstarp mesons, presented in Section~\ref{sec:results}.
Subsequently, the so-obtained correlation functions are mapped into the ones of ($\Dplus\leftarrow\Dstarp$)$\uppi$ and ($\Dplus\leftarrow\Dstarp$)K pairs, respectively. The transformation of the momentum basis is performed using GENBOD phase-space simulations~\cite{James:1968gu} of the $\Dstarpm \rightarrow \mathrm{D}^\pm \uppi^{0}$ decay, as in this case the kinematics are most stringently constrained. Contributions to the \Dplus-meson yield from decays of other excited charm resonances are considered to be negligible~\cite{ParticleDataGroup:2022pth}.

A flat correlation function is assumed for sources of background that are not expected to lead to correlations, or that can be assumed negligible due to their small $\lambda$ scaling parameter. They include contributions from particle pairs involving non-primary light-flavor mesons and contamination of the samples, as well as non-prompt \DorDstarp mesons. Especially, the correlation of primary light-flavor mesons with non-prompt \Dplus mesons is studied in analogy to \Dplus mesons from \Dstar decays, assuming Coulomb-only interaction, as it is associated to a non-negligible $\lambda$ parameter of $\sim5$\%.
The decay kinematics for $\mathrm{B}^+ \to \Dplus+\mathrm{X}$ decay is simulated and the correlation function of B mesons and light-flavor meson pairs is mapped into the one of the daughter \Dplus and light-flavor mesons. As the phase space available for the decay is much larger compared to the $\Dstarp\to\Dplus$ case, the information on the interaction between beauty and light-flavor hadrons is lost, leading to a flat correlation.

In total, four(three) contributions to $C_\text{femto} (\kstar)$ of the \DorDstarNoSignK and \DorDstarNoSignPi systems can be identified. The individual $\lambda_\mathrm{ij}$ parameters are combined, based on how the corresponding correlation functions are obtained: $\lambda_\mathrm{gen}$ is associated with the correlation function obtained from primary signal particles only, $\lambda_\mathrm{SB}$ to the one from \DorDstarp-meson background candidates, $\lambda_\mathrm{\mathrm{D} \leftarrow \Dstar}$ to the one obtained using \Dplus mesons from \Dstarp-meson decays, and $\lambda_\mathrm{flat}$ contains all other femtoscopic contributions. The combined $\lambda$ parameters for each system can be found in Table~\ref{tab:lambdapars}. 

\subsection{Residual contributions}
\label{subsec:residual}
Energy--momentum conservation effects and the production of particles within jet-like structures introduce an enhancement of the correlation function and represent a residual background $C_\text{non-femto} (\kstar)$ not related to FSI and already introduced in Eq.~\ref{eq:rawcf}, which has to be taken into account. The contribution of jet-like structures was observed in several meson--meson~\cite{Abelev:2012151,Abelev:2012sq,ALICE:2018ysd, ALICE:2011kmy, ALICE:2015hav}, meson--baryon~\cite{ALICE:2021cpv,ALICE:2019gcn}, and baryon--antibaryon~\cite{ALICE:2021cyj} femtoscopic analyses. They are related to initial hard processes at the parton level~\cite{ALICE:2016jjg}, and not to femtoscopic FSI.
The correlation function used to model the residual background, $C_\text{non-femto} (\kstar)$, is obtained from MC simulations, where the FSI is absent. It is further multiplied by a constant $N$, which is a free parameter and accounts for a possible bias due to the chosen normalization region of the raw data. In the case of the \DPi and \DstarK systems, an additional polynomial of the form $p(\kstar)=a{\kstar}^2$ and $p(\kstar) = a \kstar$, respectively, are added to the MC correlation function $C_\text{MC}(\kstar)$ to better fit the background model to $C_\text{raw}(\kstar)$ at intermediate \kstar. This introduces an additional free parameter $a$ and leads to the following expression for the residual background $C_\text{non-femto} (\kstar)=N\times[C_\text{MC}(\kstar)+p(\kstar)]$.

\subsection{Modeling of the correlation function}

In order to extract the unknown $C_{\rm{gen}}(\kstar)$ from the raw data, which is needed to study the residual strong interaction between the different particle pairs of interest, a model is built according to Eq.~\ref{eq:rawcf}, taking into account all the relevant background contributions discussed in the previous sections. 

In the case of \DPi and \DK pairs, all the sources of background, mentioned and explained in detail in Sections~\ref{subsec:FSI} and~\ref{subsec:residual}, are present. Therefore, the model takes the form
\begin{equation}
\label{eq:allContribDKPi}
\begin{split}
C_\mathrm{raw}(\kstar) = \lambda_\mathrm{SB} C_\mathrm{SB}(\kstar) + C_\text{non-femto}(k^*) \left[ \lambda_\mathrm{gen} C_\mathrm{gen}(\kstar) + \lambda_{\mathrm{D} \leftarrow \Dstar}C_{\mathrm{D} \leftarrow \Dstar} (\kstar) +  \lambda_\mathrm{flat}\right],
\end{split}
\end{equation}
where $C_\mathrm{SB}(\kstar)$ is the correlation function arising from the \Dplus-meson combinatorial background,\break $C_\text{non-femto}(\kstar)$ is the correlation function that describes the residual correlation not associated to FSI and mainly coming from jet-like contributions, and $C_\mathrm{D\leftarrow\Dstar}(\kstar)$ is the correlation function associated to the \Dplus mesons from \Dstarp decays. Finally, $\lambda_\mathrm{flat}$ accounts for all femtoscopic background contributions, assumed to be flat.
Notably, as $C_\mathrm{SB}(\kstar)$ is obtained in a data-driven approach, it already includes possible residual jet-like contributions and thus does not have to be multiplied by $C_\text{non-femto}(\kstar)$.

The model for the \DstarK correlation functions is similar to the one used for \DPi and \DK correlations, with the difference, that the contribution from excited charm states is assumed to be negligible, hence $\lambda_{\mathrm{D} \leftarrow \Dstar}=0$. The same assumption holds for \DstarPi correlation functions. In this case, however, the combinatorial background is already subtracted using the sideband-less approach described by Eq.~\ref{eq:sblessapproach} in Section~\ref{subsec:FSI}. Therefore, the final model is given by Eq.~\ref{eq:allContribDKPi}, with $\lambda_{\mathrm{D} \leftarrow \Dstar}=0$ and $\lambda_\mathrm{SB}=0$.

\begin{figure}[!tb]
    \centering
        \includegraphics[width=0.49\linewidth]{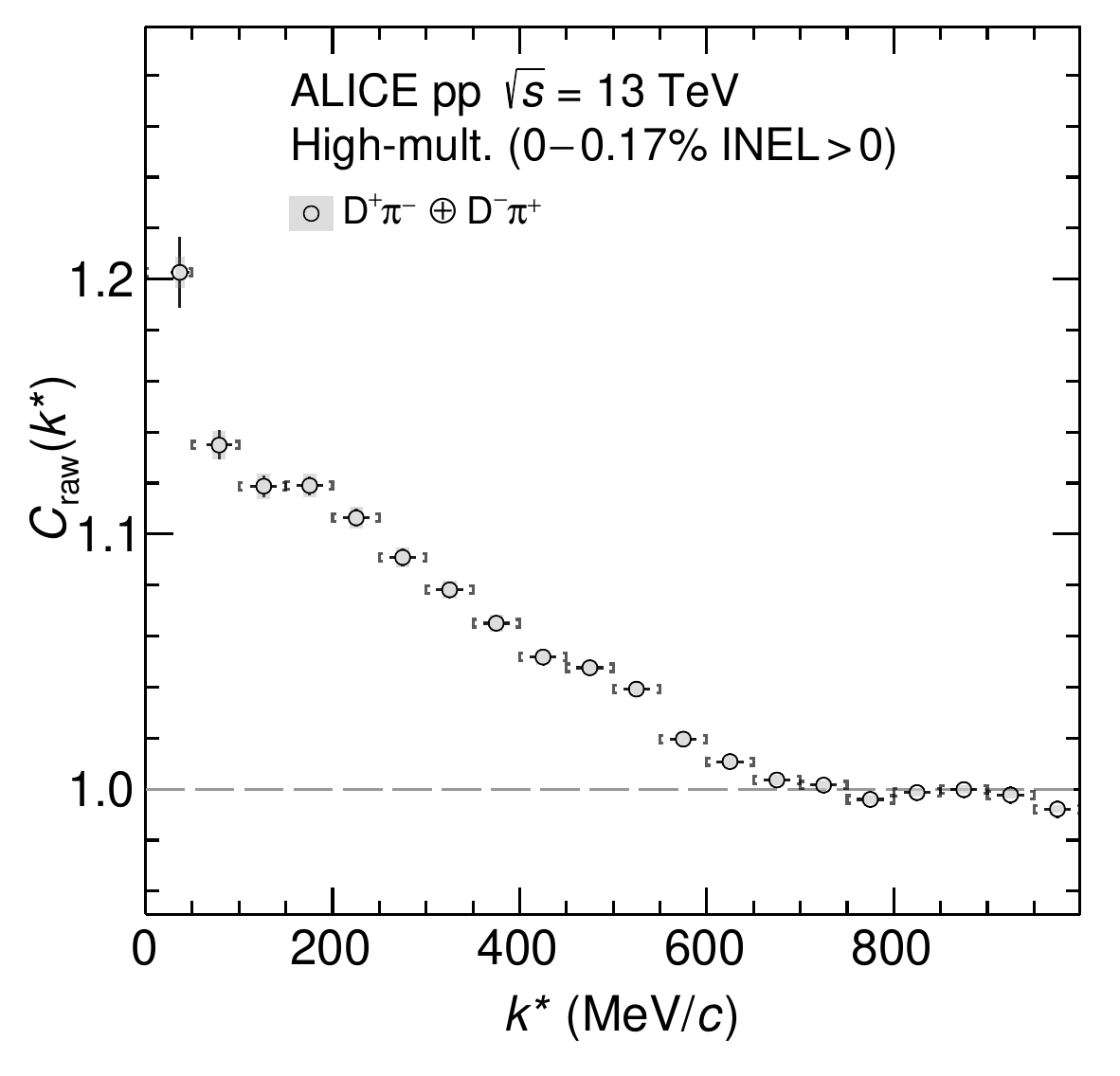}
        \includegraphics[width=0.49\linewidth]{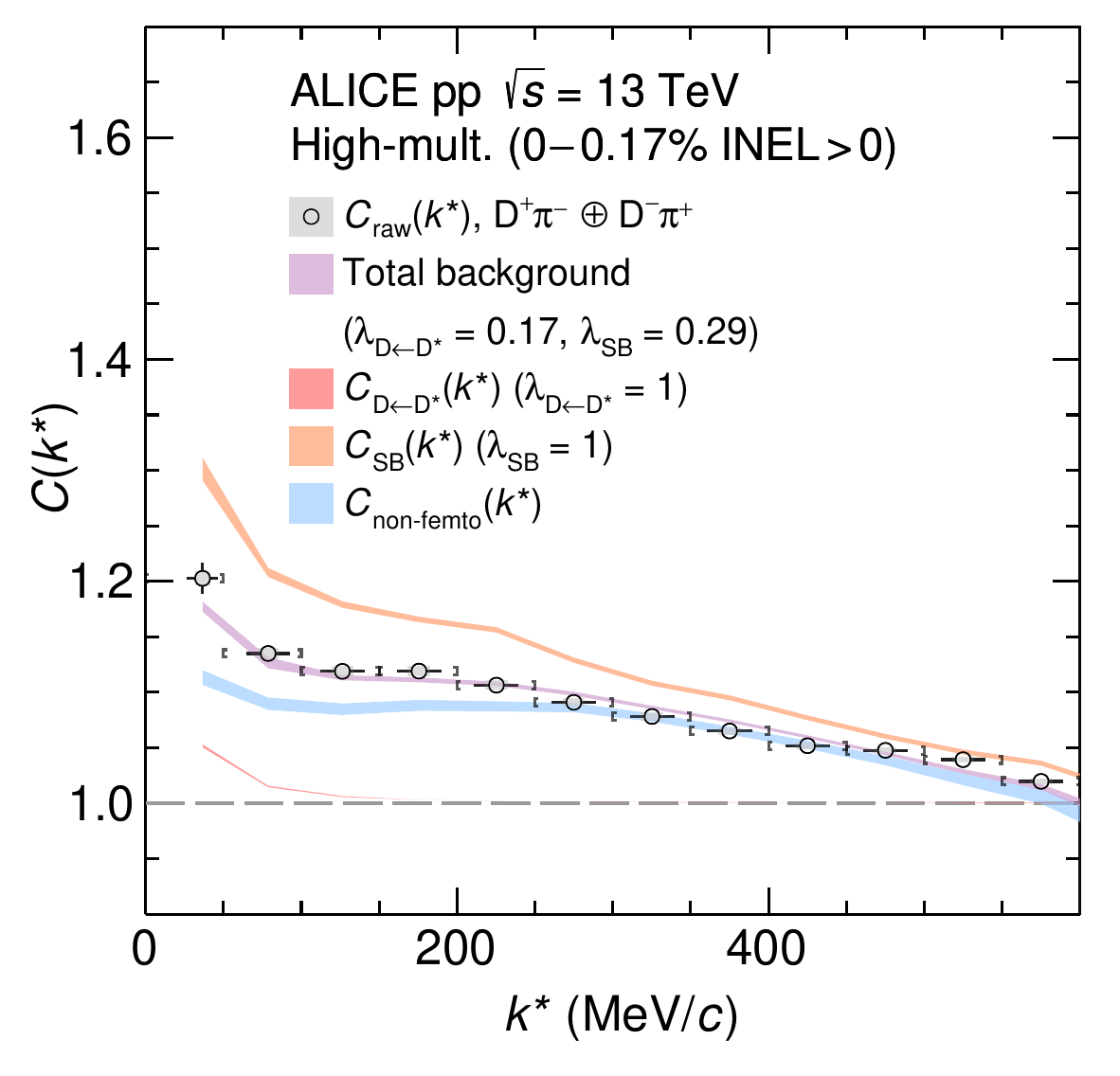}
    \includegraphics[width=0.49\linewidth]{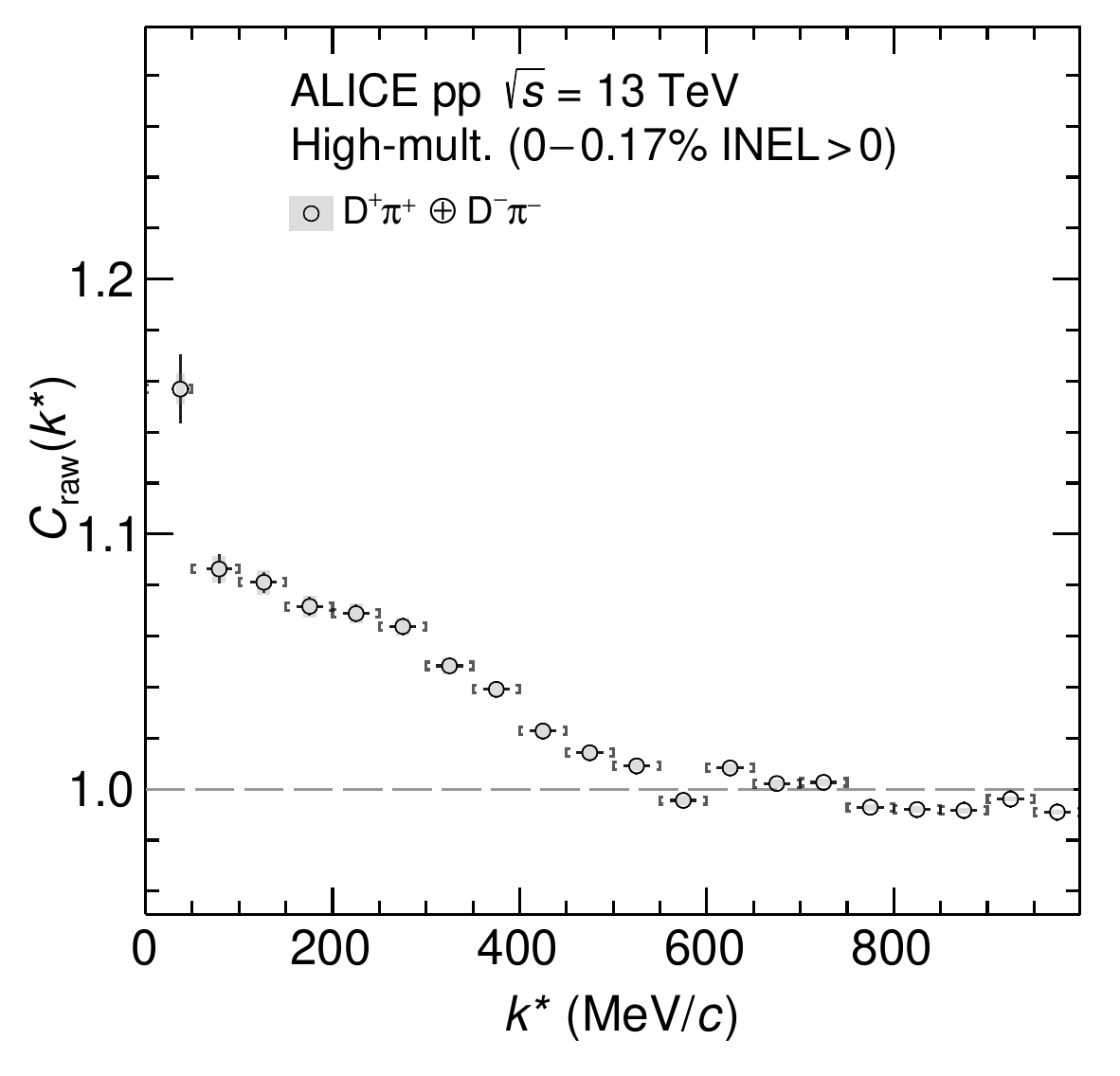}
        \includegraphics[width=0.49\linewidth]{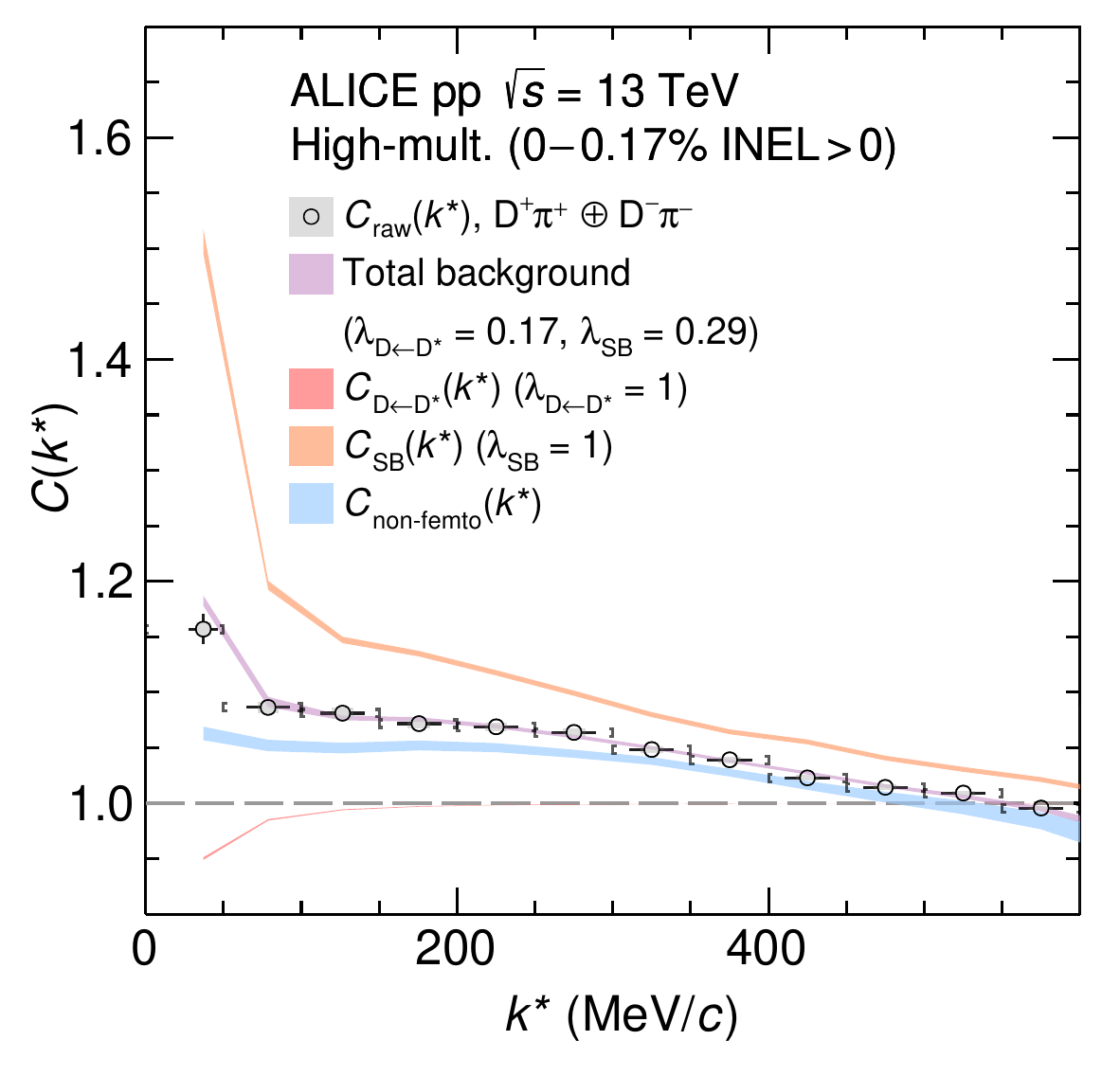}
     \caption{Experimental \DPi raw correlation functions ($C_\mathrm{raw}(\kstar)$) with statistical (bars) and systematic uncertainties (boxes) (left column), and background contributions to the experimental correlation functions (right column). The width of the bands corresponds to the total uncertainty $\sigma_\mathrm{tot}=\sqrt{\sigma_\mathrm{stat}^2+\sigma_\mathrm{syst}^2}$. The violet band describes the total background, fitted to the data, and used to extract the genuine correlation function from the raw signal. This band consists of several contributions, which are also shown individually in the figure, scaled by the appropriate $\lambda$ parameter. The results are shown for opposite-charge (first row) and same-charge (second row) pairs.}
    \label{fig:bkg_pions}
\end{figure}

\begin{figure}[!tb]
    \centering
    \includegraphics[width=0.49\linewidth]{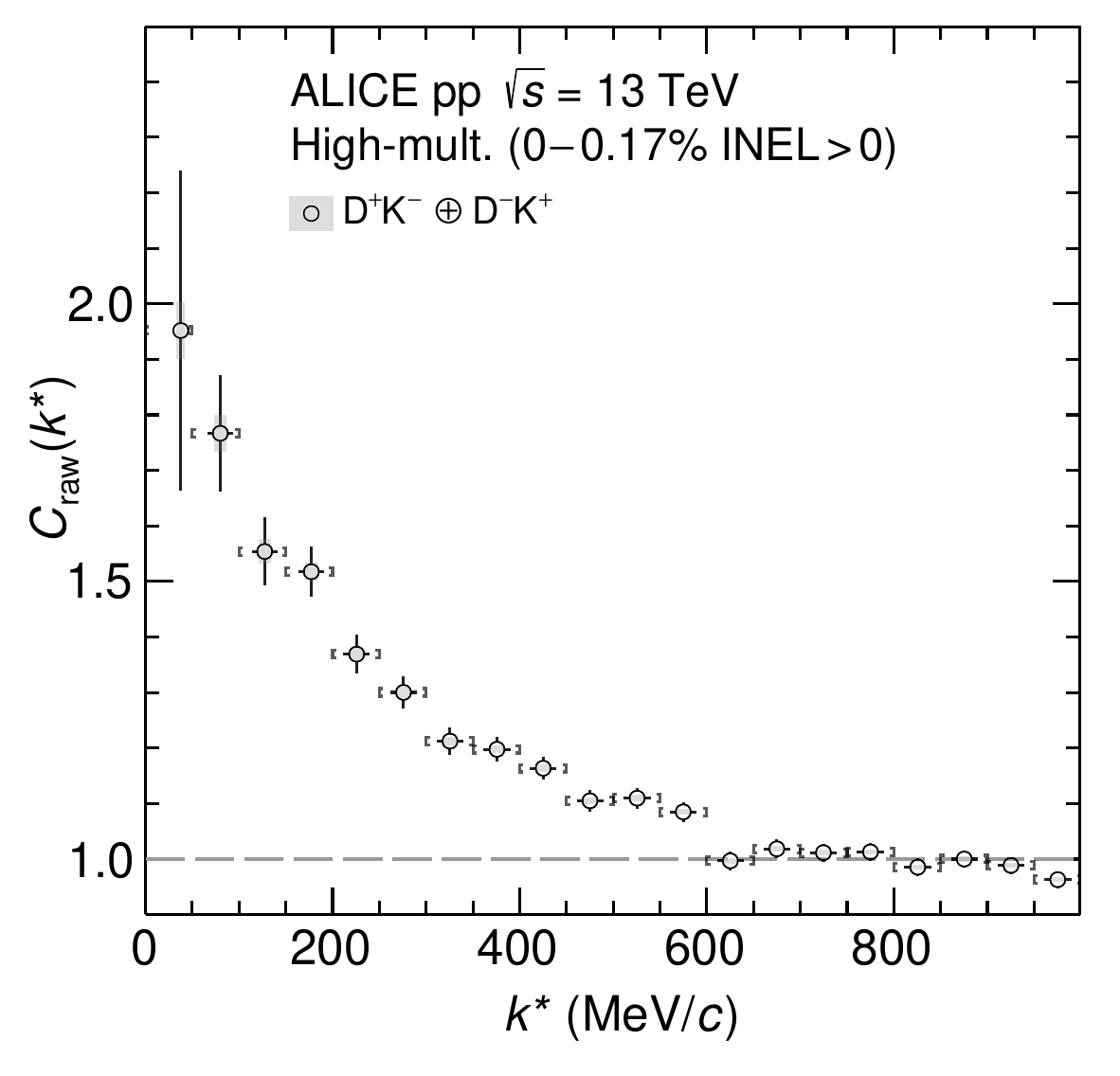}
        \includegraphics[width=0.49\linewidth]{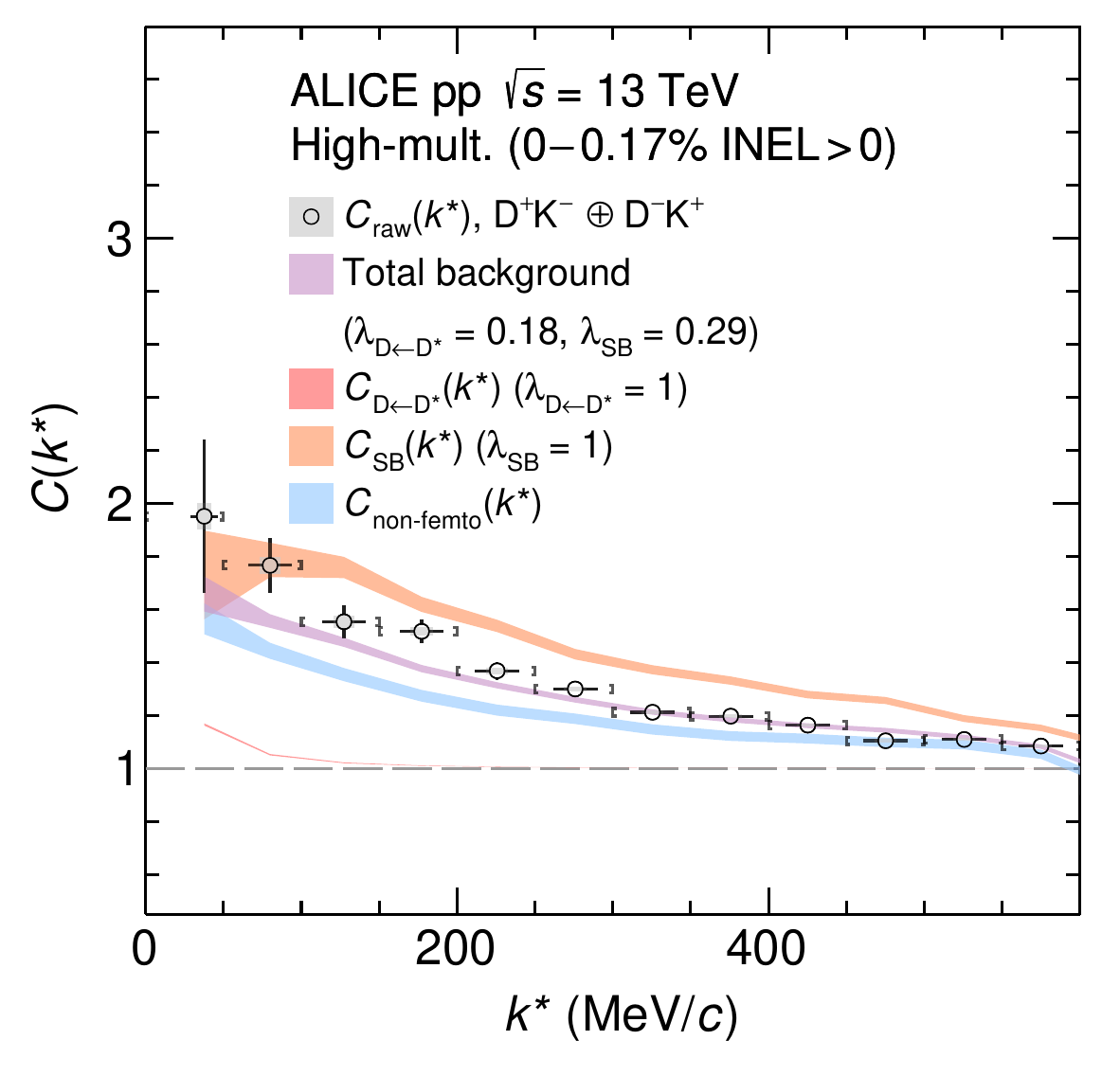}
    \includegraphics[width=0.49\linewidth]{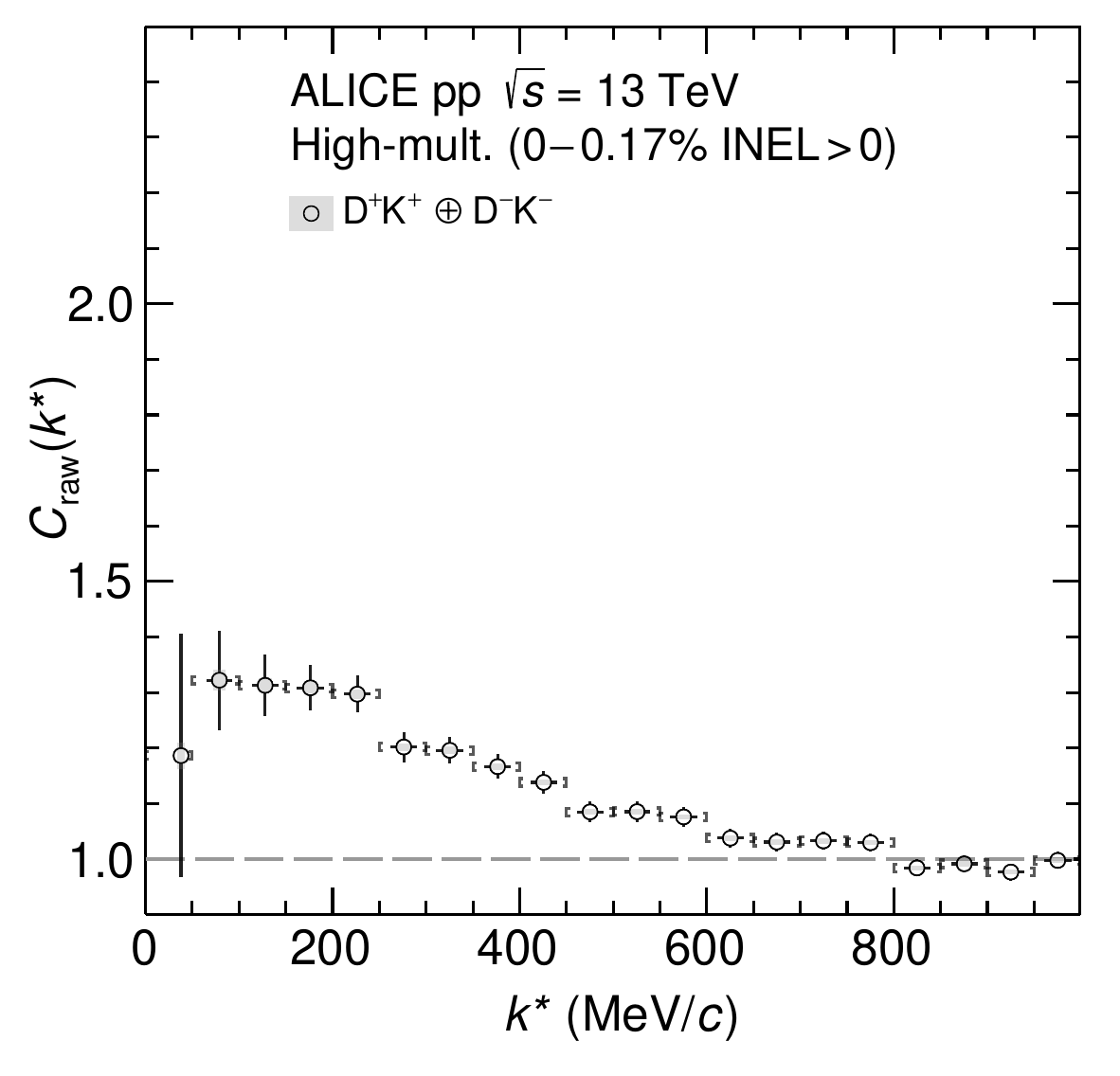}
        \includegraphics[width=0.49\linewidth]{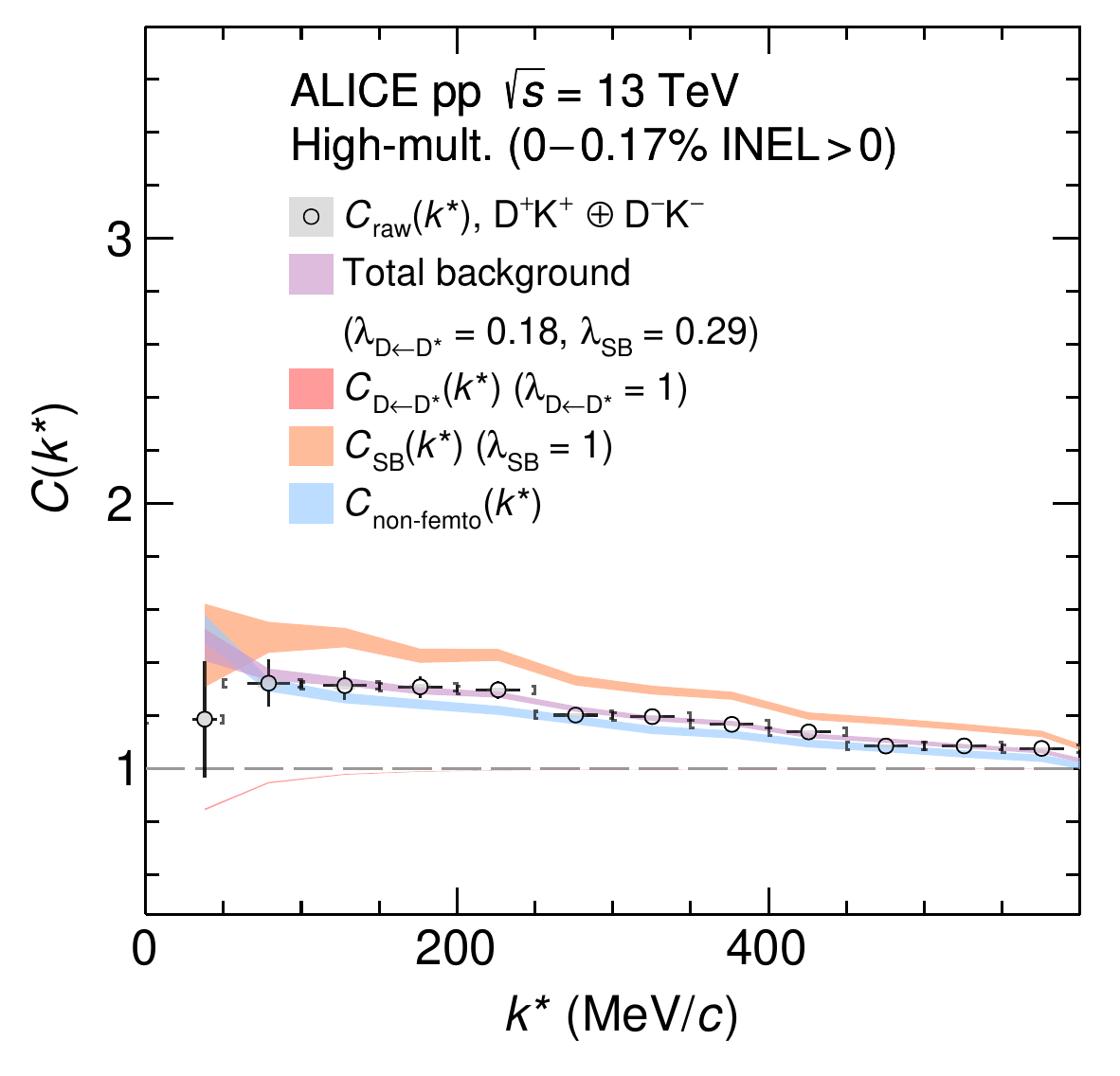}
     \caption{Experimental DK raw correlation functions ($C_\mathrm{raw}(\kstar)$) with statistical (bars) and systematic uncertainties (boxes) (left column), and background contributions to the experimental correlation functions (right column). The width of the bands corresponds to the total uncertainty $\sigma_\mathrm{tot}=\sqrt{\sigma_\mathrm{stat}^2+\sigma_\mathrm{syst}^2}$. The violet band describes the total background, fitted to the data, and used to extract the genuine correlation function from the raw signal. This band consists of several contributions, which are also shown individually in the figure, scaled by the appropriate $\lambda$ parameter. The results are shown for opposite-charge (first row) and same-charge (second row) pairs.}
    \label{fig:bkg_kaons}
\end{figure}

To determine the free parameters related to $C_\text{non-femto}(\kstar)$, a background model is defined by imposing $C_\mathrm{gen}(\kstar) = 1$ in Eq.~\ref{eq:allContribDKPi} for all pair combinations.  
The resulting expressions are fitted directly to the raw data in the range of $\kstar\in[100,600]~\MeVc$ for \DPi correlations and $\kstar\in[200,400]~\MeVc$ for \DK. The chosen fit range for correlations involving \DstarPi is $\kstar\in[300,1000]~\MeVc$, while it is $\kstar\in[250,500]~\MeVc$ for \DstarK. The fit ranges are tuned to select a \kstar region in which the femtoscopic correlations are expected to be negligible.
The different sources of background, together with the total background model (violet band) and the raw data, are reported in the right panels of Figs.~\ref{fig:bkg_pions} and~\ref{fig:bkg_kaons} for both the \DPi and \DK correlation functions, respectively. The blue band represents the residual $C_\text{non-femto}(\kstar)$, the orange band the combinatorial background $C_\mathrm{SB}(\kstar)$, and the red band the contribution arising from the feed-down of \Dstarp to \Dplus, $C_{\mathrm{D} \leftarrow \Dstar} (\kstar)$.
Once the parameters are fixed from the fit, the genuine correlation function $C_\mathrm{gen}(\kstar)$ is extracted from the raw data via Eq.~\ref{eq:allContribDKPi}, adapted to the pair of interest.

\section{Systematic uncertainties} \label{sec:syst}
The genuine correlation functions, which are the observables used to extract information on the residual strong final state interaction, are affected by several sources of systematic uncertainty related to the selection criteria or the background corrections to the raw data. Such uncertainty contributes to the systematic uncertainty of the scattering parameters, together with the systematic uncertainties associated with the source parameters and the choice of the fit range. The details on how the systematic uncertainties are estimated are provided in the following paragraphs.

\subsection{Genuine correlation function}
The choice of the selection criteria of the light-flavor and \DorDstarNoSign-meson candidates determines the single-particle properties of the sample and hence the distributions of the pairs in the same (mixed) events. Therefore, an impact on the raw correlation function is expected, which is then propagated to the genuine correlation function.
To estimate the systematic uncertainty associated with this contribution, the selection criteria mentioned in Section~\ref{sec:analysis} are varied, and the raw correlation functions are recomputed for each set of variations. On the raw correlation function, the relative systematic uncertainty is below $3\%$ in the case of \DorDstarNoSignK pairs and below $1\%$ in the case of \DorDstarNoSignPi pairs.

The uncertainties on the $\lambda$ parameters, which affect the modeling of the background, are dominated by the uncertainty on the fraction of non-prompt charm mesons, \Dplus mesons from \Dstarp decays, and light-flavor mesons from the decay of long-lived resonances as well as the purity of the \DorDstarp meson candidates.
It is estimated by varying the fractions of charm mesons according to the uncertainties stated in Section~\ref{sec:analysis}. The fractions of strongly-decaying long-lived resonances feeding into the light-flavor mesons, which are estimated using \fist, are varied by 10\%~\cite{KorwieserSource} and the purity of \DorDstarp mesons by 2\%~\cite{ALICE:2022enj}. This leads to a variation of $\sim10\%$ of the $\lambda$-parameter values.\\
The systematic uncertainty on the background model is estimated by propagating the systematic uncertainties of the raw correlation functions and by varying the fractions according to the uncertainties stated above. Additionally, the fit range of the background model is varied in order to account for possible systematic effects related to the fit procedure. For particle pairs involving \Dplus mesons, where the feed-down contribution from \Dstarp decays is modeled assuming Coulomb-only interaction, the uncertainty on the effective source parameterization, reported in Table~\ref{tab:effrad}, represents an additional source of systematic uncertainty of the background model.

The total systematic uncertainty of the genuine correlation functions, computed taking into account all the contributions mentioned above, is found to be below 1\% for opposite-charge \DPi, below 2\% for same-charge \DPi, below 10\% for same-charge \DK, below 15\% for opposite-charge \DK, below 2.5\% for \DstarPi, below 7\% for same-charge \DstarK, and below 25\% for opposite-charge \DstarK. In the low \kstar region, the correlation functions are the most affected by the systematic uncertainties. The larger relative systematic uncertainty of the \DorDstarNoSignK correlation functions with respect to the ones of \DorDstarNoSignPi arises from the propagated uncertainty of the raw correlation functions, which is related to the light-flavor meson selections. Overall, this represents the main source of systematic uncertainty of the genuine correlation functions, followed by the uncertainty on the $\lambda$ parameters.

\subsection{Scattering lengths}
The systematic uncertainty associated to the extraction of the scattering lengths, discussed in the next section, besides the one related to the genuine correlation functions, is obtained by taking into account the choice of the fit range and the lack of precise knowledge of the source function. The first is estimated by varying the fit range by 50~$\MeVc$, and the second one by performing the fit with different effective source parameters, determined according to the uncertainties reported in Table~\ref{tab:effrad}. The latter represents the largest contribution to the systematic uncertainties on the scattering lengths, besides the propagated systematic uncertainties of the genuine correlation functions.

\section{Results}
\label{sec:results}

\begin{table}[!tb]
\caption{Scattering lengths of the available theoretical models for the \DPi interactions. The values are reported separately for the different isospin states.}
\centering
\renewcommand*{\arraystretch}{1.2}
\scalebox{1}{
\begin{tabular}{l c | c c}
Model & & \multicolumn{2}{c}{$a_0~(\fm)$}\\
\hline\hline                                                     &           & $\DPi(I=\sfrac{3}{2})$              & $\DPi(I=\sfrac{1}{2})$          \\\cline{3-4}
L. Liu \emph{et al.}~\cite{Liu:2012zya}                         &           & $-0.100\pm0.002$              & $0.37^{+0.03}_{-0.02}$    \\
X. Y. Guo \emph{et al.}~\cite{Guo:2018kno}                      &           & $-0.11$                       & 0.33                      \\
\multirow{2}{*}{Z. H. Guo \emph{et al.} ~\cite{Guo:2018tjx}}    & Fit-1B    & $-0.101^{+0.005}_{-0.003}$    & $0.31^{+0.01}_{-0.01}$    \\
                                                                & Fit-2B    & $-0.099^{+0.003}_{-0.004}$    & $0.34^{+0.00}_{-0.03}$    \\
B. L. Huang \emph{et al.}~\cite{Huang:2021fdt}                  &           & $-0.06\pm 0.02$               & $0.61\pm0.11$             \\
\multicolumn{2}{l|}{J. M. Torres-Rincon \emph{et al.}~\cite{Torres-Rincon:2023qll}}           & $-0.101$                      & 0.423                     \\
\cline{3-4}
& & $\DstarPi(I=3/2)$ & $\DstarPi(I=1/2)$\\
\cline{3-4}
Z.-W. Liu~\cite{Liu:2011mi}& & $-0.13-0.00036i$&$0.27-0.00036i$ \\

\hline\hline
\end{tabular}
}
\label{tab:scatParamPi}
\end{table}

\begin{table}[!b]
\caption{Scattering lengths of the available theoretical models for the \DK interactions. The values are reported separately for the different strangeness and isospin states. The real and imaginary components are associated with elastic and inelastic processes, respectively.}
\centering
\renewcommand*{\arraystretch}{1.2}
\scalebox{1}{
\begin{tabular}{l c | c | c c}
Model & & \multicolumn{3}{c}{$a_0~(\fm)$}\\
\hline\hline                                                     &           & $\DK(I=1)$                                            & $\DAntiK(I=1)$                     & $\DAntiK(I=0)$\\
\cline{3-5} 
L. Liu \emph{et al.}~\cite{Liu:2012zya}                         &           & $0.07\pm0.03 +  0.17^{+0.02} _{-0.01}i$           & $-0.20\pm0.01$            & $0.84^{+0.17}_{-0.22}$\\
X. Y. Guo \emph{et al.}~\cite{Guo:2018kno}                      &           & $-4.87  \times 10^{-2}$                            & $-0.22$                   & 0.46\\
\multirow{2}{*}{Z. H. Guo \emph{et al.} ~\cite{Guo:2018tjx}}    & Fit-1B    & $0.06^{+0.05}_{-0.03} + 0.30 ^{+0.09}_{-0.05} i$  & $-0.18^{+0.01}_{-0.01}$   & $0.96^{+1.44}_{-0.44}$\\
                                                                & Fit-2B    & $0.05^{+0.04}_{-0.03} + 0.17^{+0.02}_{-0.03}i$    & $-0.19^{+0.02}_{-0.02}$   & $0.68^{+0.17}_{-0.16}$\\
B. L. Huang \emph{et al.}~\cite{Huang:2021fdt}                  &           & $-0.01\pm0.03$                                    & $-0.24\pm0.02$           & $1.81\pm0.48$ \\
\multicolumn{2}{l|}{J. M. Torres-Rincon \emph{et al.}~\cite{Torres-Rincon:2023qll}}             & $-0.027+0.083i$                                   & $-0.233$                  & $0.399$\\

\cline{3-5}  
 & &  $\DstarK(I=1)$& $\Dstar\overline{K}(I=1)$ & $\DstarK(I=0)$ \\
\cline{3-5}  
Z.-W. Liu~\cite{Liu:2011mi} & & $-0.022+0.18i$&$-0.19-\num{1.7e-6}i$ & $0.29+\num{5.2e-6}i$\\
\hline\hline
\end{tabular}
}
\label{tab:scatParamK}
\end{table}

\begin{figure}[!b]
    \centering
    \includegraphics[width=0.49\linewidth]{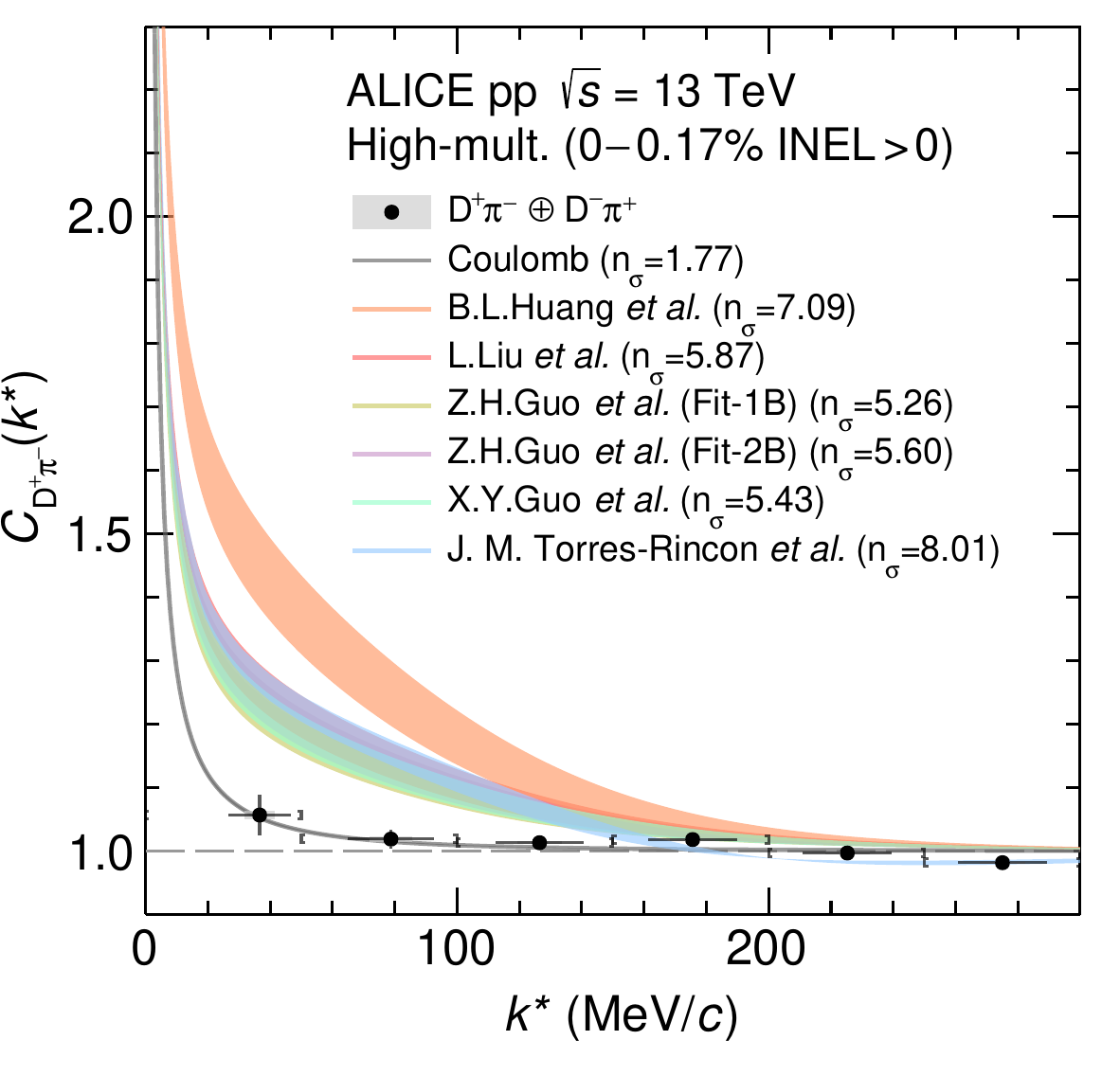}
    \includegraphics[width=0.49\linewidth]{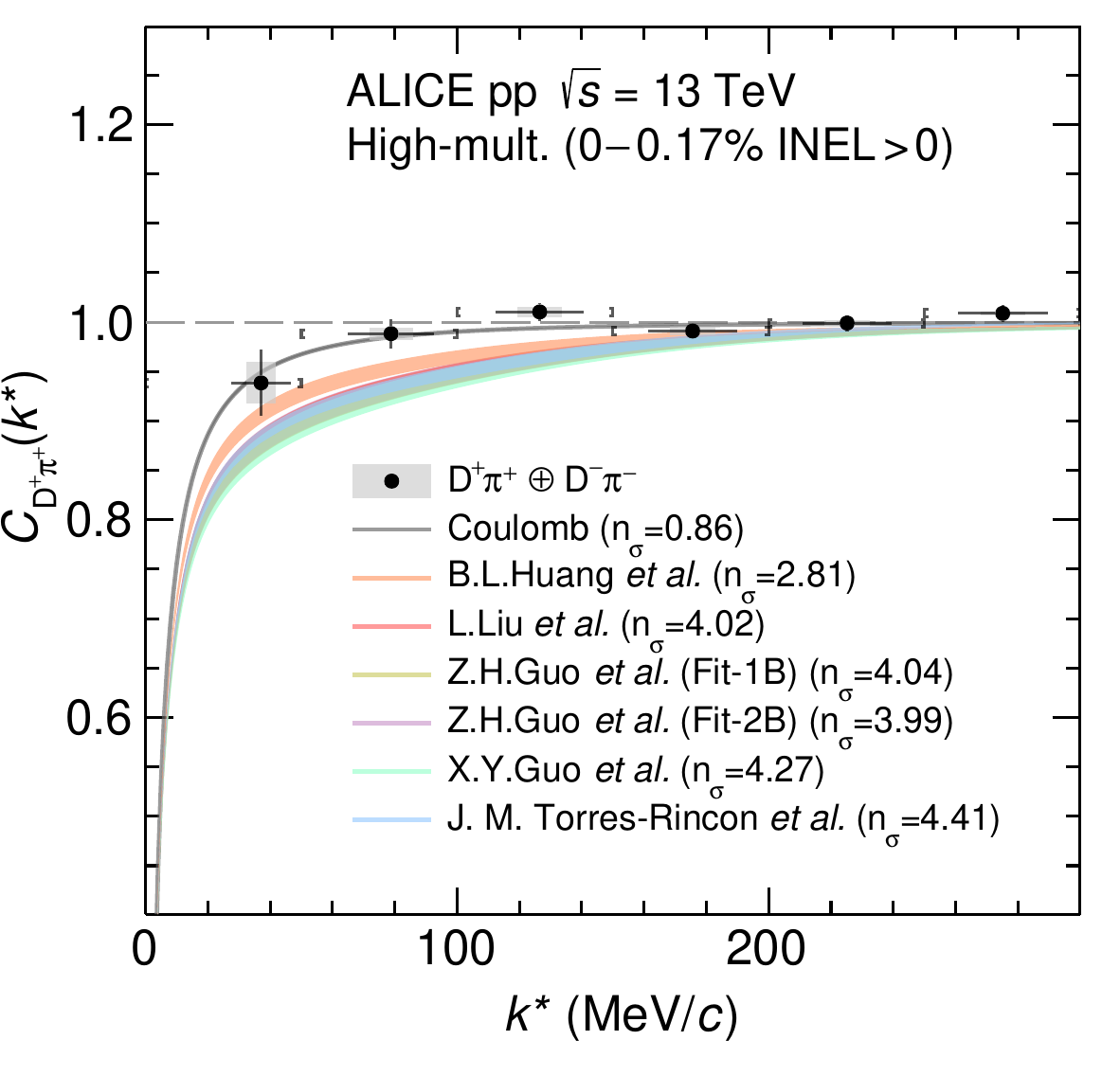}
    \includegraphics[width=0.49\linewidth]{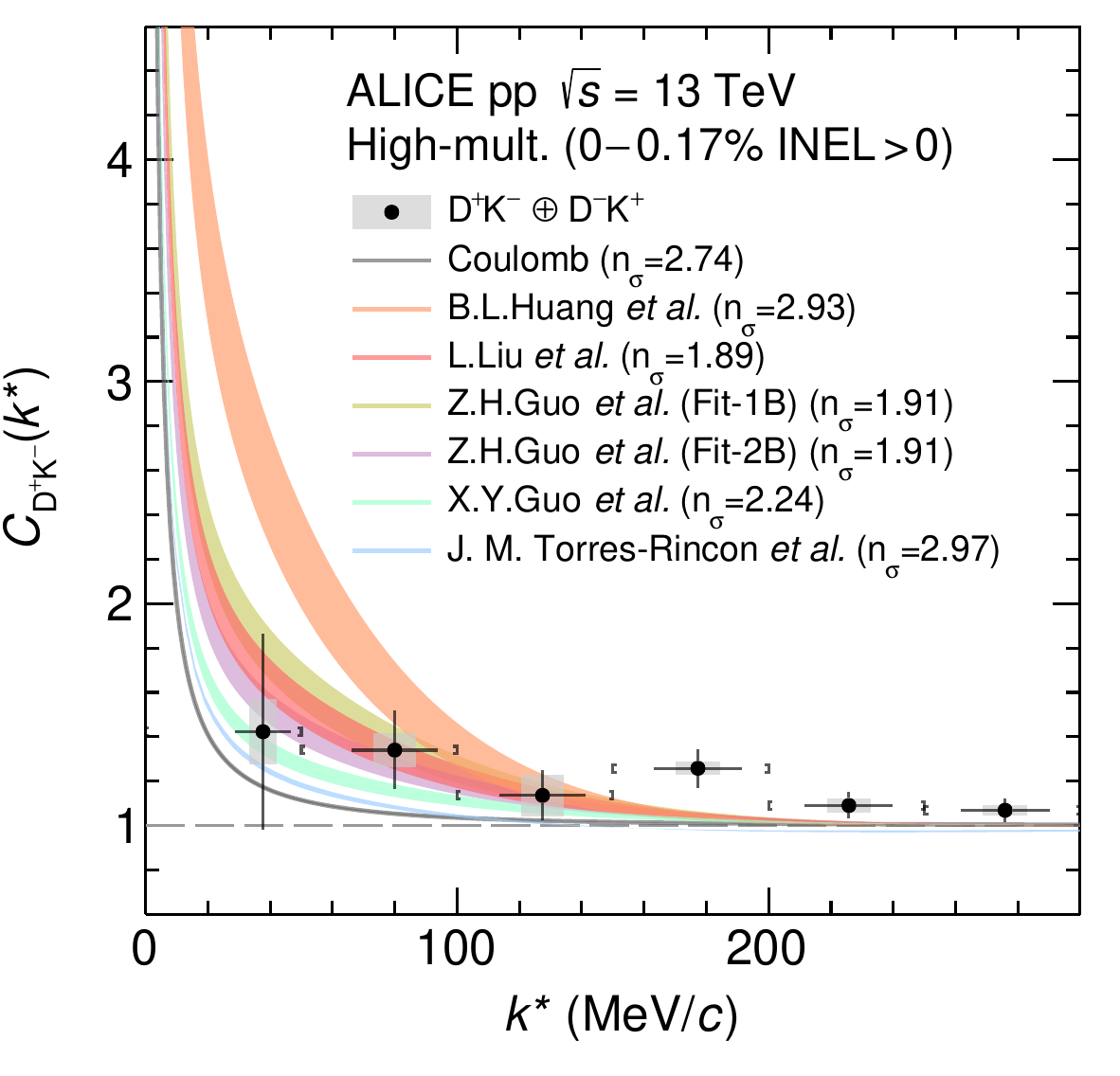}
    \includegraphics[width=0.49\linewidth]{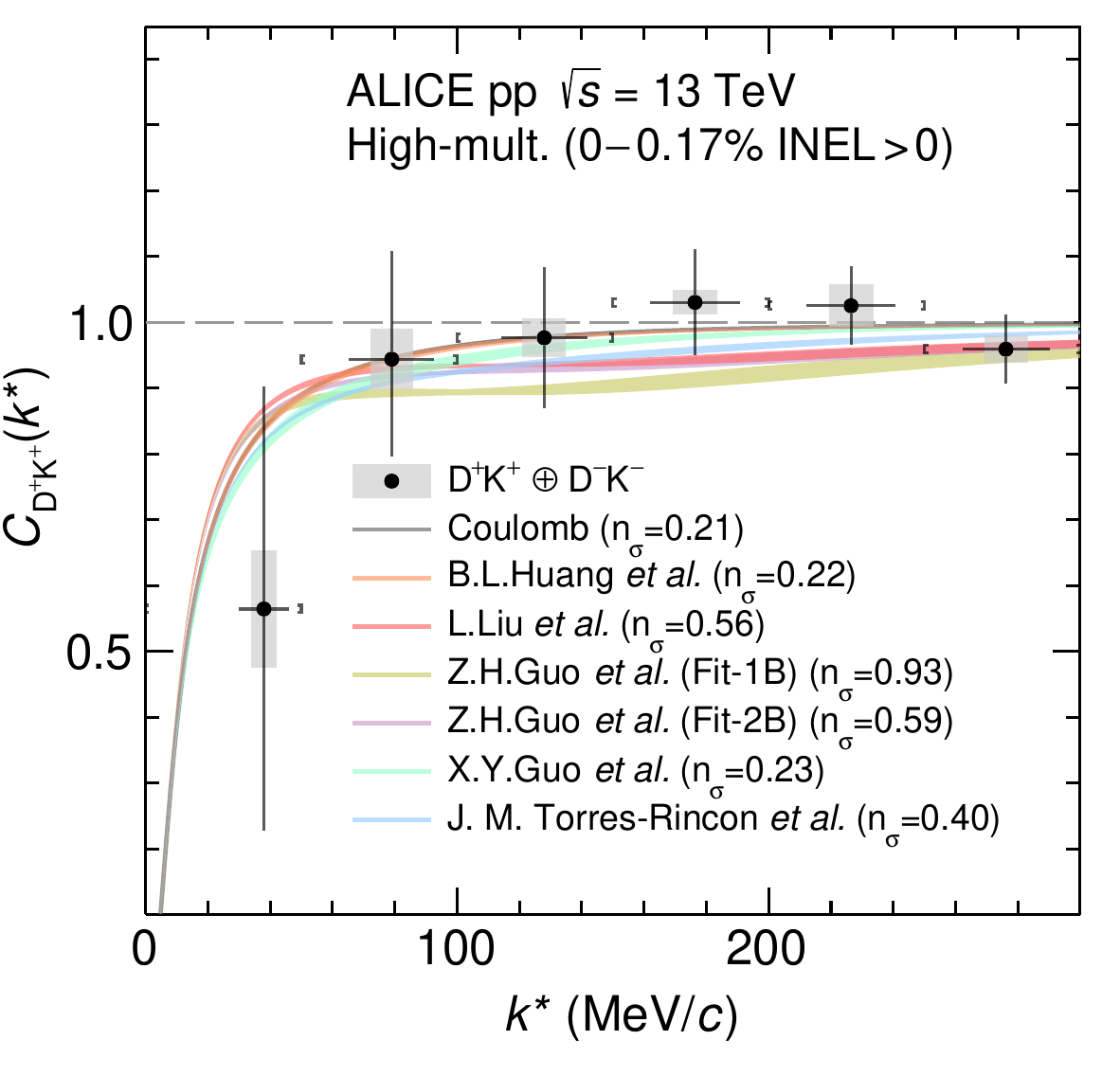}
    \caption{Genuine correlation functions with statistical (bars) and systematic uncertainties (boxes) compared to theoretical model predictions (bands), listed in Tables~\ref{tab:scatParamPi} and~\ref{tab:scatParamK}. The width of the theoretical bands represents the uncertainty related to the source. The number of standard deviations $n_{\sigma}$ is reported for each model in the legend. The results are shown for \DPi (first row) and DK (second row) for the opposite- (left column) and same-charge (right column) combinations.}
    \label{fig:modelsdplus}
\end{figure}

\begin{figure}[!t]
    \centering
    \includegraphics[width=0.49\linewidth]{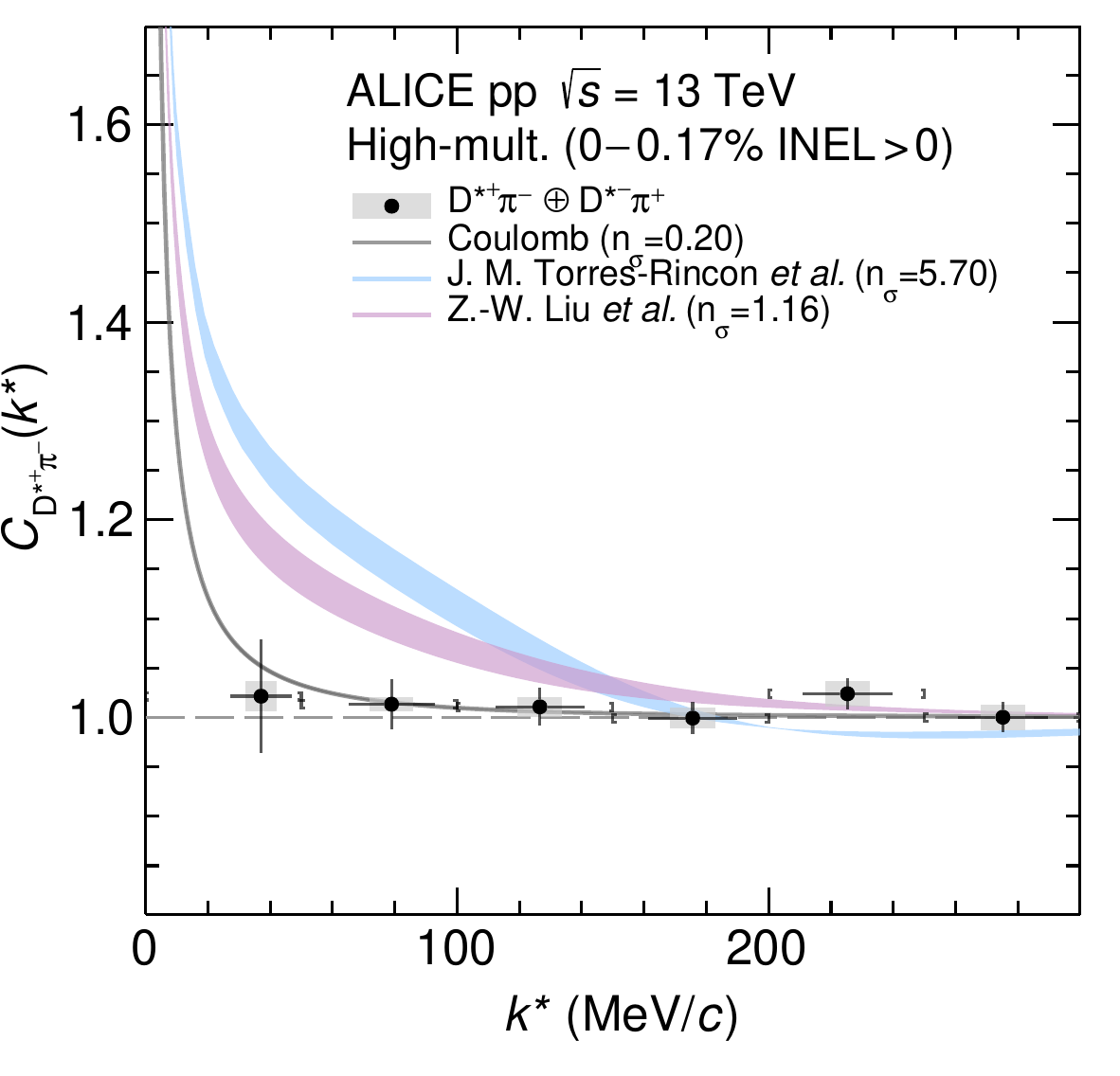}
    \includegraphics[width=0.49\linewidth]{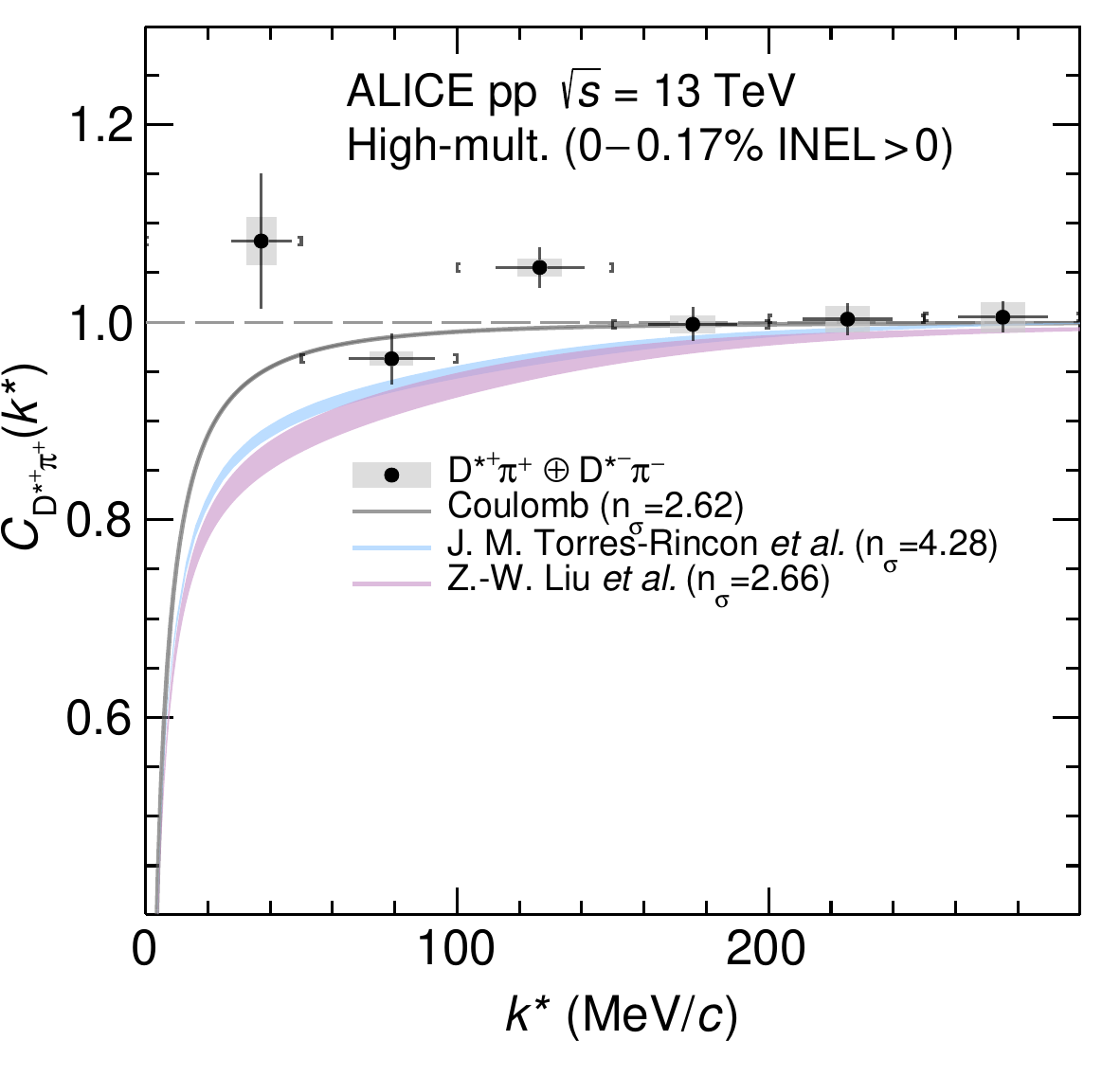}\\
    \includegraphics[width=0.49\linewidth]{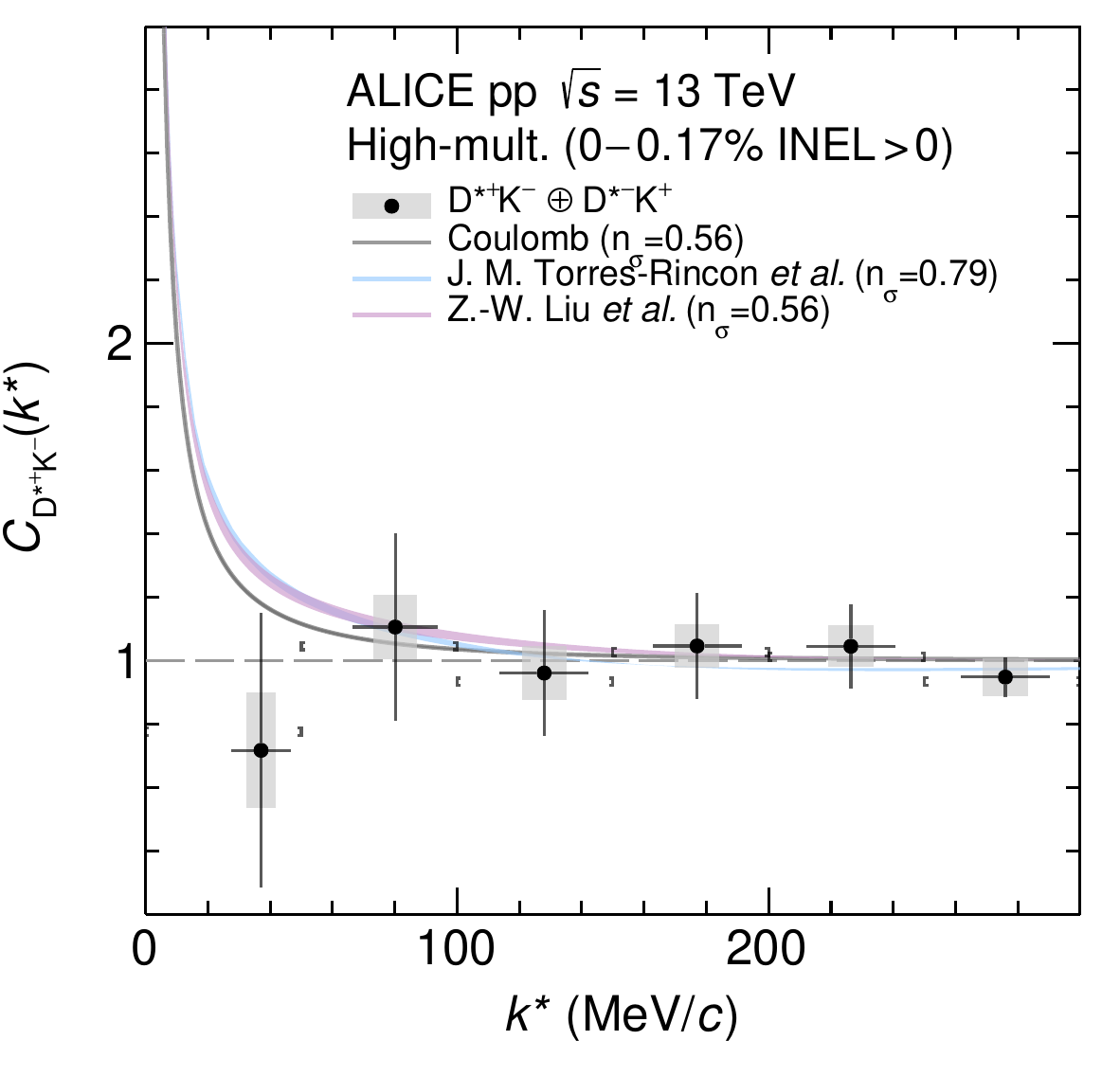}
    \includegraphics[width=0.49\linewidth]{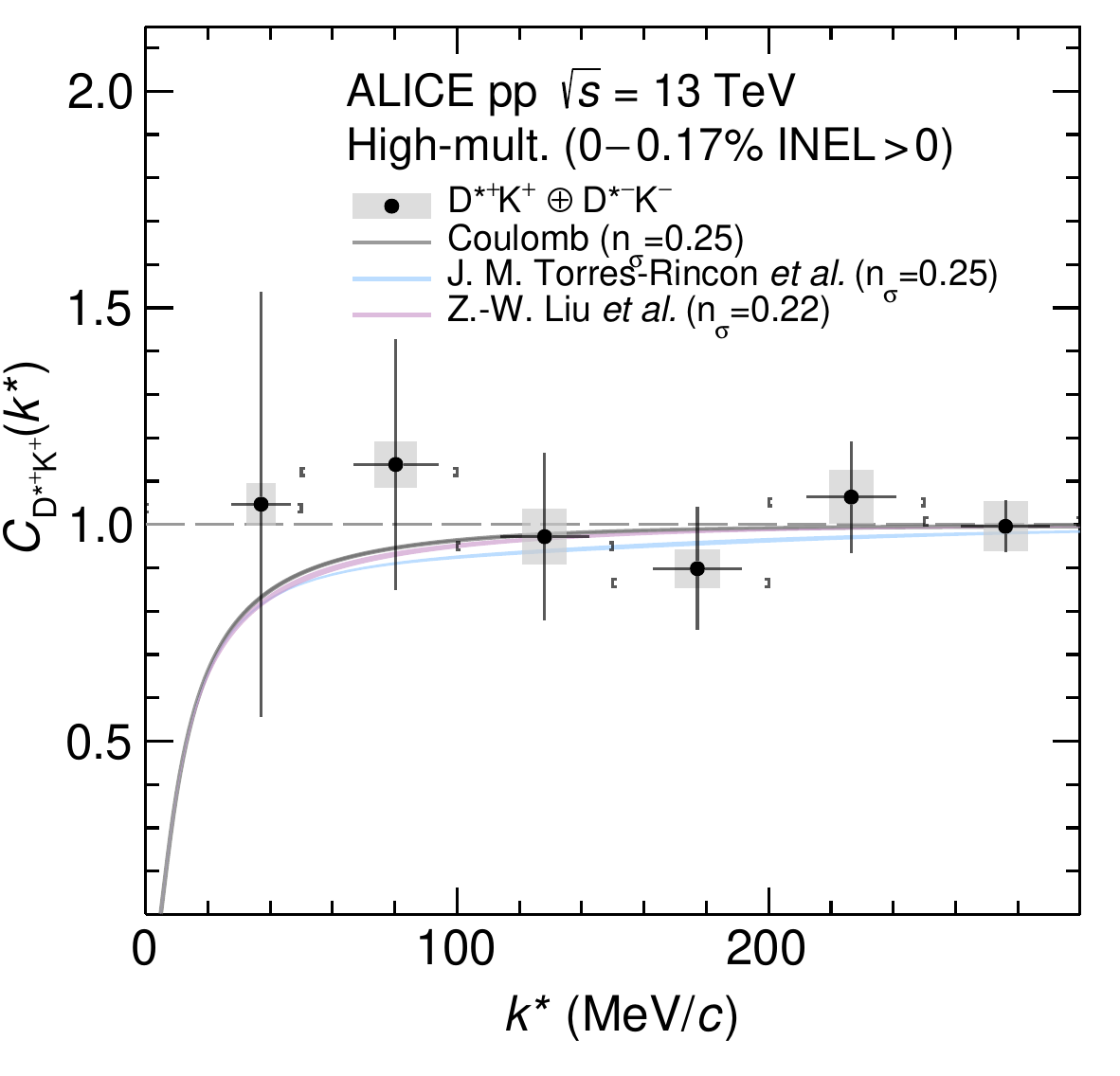}
    \caption{Genuine correlation functions with statistical (bars) and systematic uncertainties (boxes) compared to theoretical model predictions (bands). The width of the bands represents the uncertainty related to the source. The number of standard deviations $n_{\sigma}$ is reported for each model in the legend. The results are shown for \DstarPi (first row) and \DstarK (second row) for the opposite- (left column) and same-chage (right column) combinations.}
    \label{fig:modelsdstar}
\end{figure}

The measured genuine correlation functions, extracted from the raw data as described in Section~\ref{sec:cf}, are shown in Fig.~\ref{fig:modelsdplus} and Fig.~\ref{fig:modelsdstar} for correlations involving light-flavor and \Dplus or \Dstarp mesons, respectively. In the femtoscopic region $\kstar<200~\MeVc$ the genuine correlation functions are sensitive to the Coulomb and strong nuclear forces and can be compared to the corresponding calculations.\\
The strong interactions between the mesons depend on the quantum numbers of the systems and can therefore be separated into different isospin and strangeness configurations. These are namely: $\mathrm{D}^{(*)}\uppi(I=3/2,1/2, S=0)$, $\mathrm{D}^{(*)}\overline{\mathrm{K}}(I=1,0,S=-1)$, and $\mathrm{D}^{(*)}\mathrm{K}(I=0,S=+1)$. 
Several theoretical predictions are available for the \DPi and \DK scattering lengths~\cite{Liu:2012zya,Guo:2018kno,Guo:2018tjx,Huang:2021fdt,Torres-Rincon:2023qll}, while only two are present for the \DstarPi and \DstarK systems~\cite{Torres-Rincon:2023qll, Liu:2011mi}. 
The models are listed below, together with a brief description of the calculation method. The corresponding scattering lengths are summarized in Tables~\ref{tab:scatParamPi} and~\ref{tab:scatParamK}.
\begin{itemize}
    \item \textbf{L. Liu \emph{et al.}}~\cite{Liu:2012zya}: The S-wave scattering lengths $a_0^\mathrm{D\uppi}(I=3/2)$, $a_0^{\DAntiK}(I=0)$, and $a_0^{\DAntiK}(I=1)$ are calculated on the lattice using Lüschers finite volume technique. Extrapolation to the physical point is performed using unitarized chiral perturbation theory (ChPT) up to next-to-leading order (NLO), where the low-energy constants (LECs) are determined by a fit to the lattice data. The latter are exploited to predict the scattering lengths in the other isospin channels.
    \item \textbf{X. Y. Guo \emph{et al.}}~\cite{Guo:2018kno}: N$^3$LO ChPT is employed and the LECs are determined by a global fit to lattice QCD data, including the S-wave scattering length from~\cite{Liu:2012zya}. A chiral expansion scheme is applied to obtain the scattering lengths at physical pion mass.
    \item \textbf{Z. H. Guo \emph{et al.}}~\cite{Guo:2018tjx}: The scattering length between the light-flavor and charmed mesons is obtained from unitarized ChPT up to NLO. The free parameters of the theory are constrained to lattice QCD calculations of the scattering length, including~\cite{Liu:2012zya}, and the finite-volume spectra. The fit is performed on different sets of the data, denoted as Fit-1B and Fit-2B. Finally, the scattering lengths are obtained using a chiral extrapolation to the physical point. 
    \item \textbf{B. Huang \emph{et al.}}~\cite{Huang:2021fdt}: Lattice QCD calculations of the finite-volume spectra and scattering lengths, including~\cite{Liu:2012zya} are used to determine the LECs of the Lagrangian formulated within unitarized heavy-meson ChPT at N$^3$LO. The scattering lengths used in this paper are obtained from the iterated method.
    \item \textbf{J. M. Torres-Rincon \emph{et al.}}~\cite{Torres-Rincon:2023qll}: The model employs unitarized ChPT with heavy-quark symmetry considerations at NLO, in a coupled-channel basis. The LECs at NLO are taken from~\cite{Guo:2018tjx}. Prediction for the \DstarPi and \DstarK scattering lengths are provided, exploiting heavy-quark spin symmetry.
    \item \textbf{Z.-W. Liu \emph{et al.}}~\cite{Liu:2011mi}: The S-wave scattering lengths of interactions involving the heavy vector meson \Dstar are derived within the framework of heavy-meson ChPT at N$^2$LO. The LECs at NLO are obtained from the mass splitting between heavy mesons and the resonance saturation model, while most of the N$^2$LO LECs are assumed to be negligible. 
 
\end{itemize}

The theoretical curves of the J.M. Torres-Rincon \emph{et al.} model were provided in a private communication with the authors and use the effective source parameterization $S_\mathrm{eff}(\rstar)$ described in Section~\ref{subsec:FSI}, with values listed in Table~\ref{tab:effrad}.
In the other cases, the scattering lengths predicted by the models are converted into correlation functions by employing Eq.~\ref{eq:CFsourcewf} with the effective source parameterization $S_\mathrm{eff}(\rstar)$. The wave function is obtained by taking into account both the Coulomb and strong interaction. The former is modeled using the well-understood Coulomb potential, while the latter is parameterized with a Gaussian potential of the form
\begin{equation}
    V(r)=V_0\exp{(-m_\uprho^2r^2)},
    \label{eq:gaussianpot}
\end{equation}
where $V_0$ is the potential strength and $m_{\uprho}$ is the mass of the lightest exchangeable meson, the $\uprho$ meson, which is the parameter that controls the potential range. The strength $V_0$ is tuned to reproduce the scattering lengths of the model~\cite{Kamiya:2022thy}. 

The theoretical models provide the scattering parameters in the (strangeness, isospin) basis, but in the experiment, the interactions are accessible only in the charge basis. 
The same-charge pairs consist of a pure isospin state. 
The opposite-charge pairs are a mixture of two isospin states, which can be addressed by solving the coupled-channel Schr\"{o}dinger equation with two isospin interaction components. 
In the case of \DorDstarNoSignPi pairs, the isospin channel $I=3/2$ is shared between the same- and opposite-charge configurations, as both have no net strangeness. 

\begin{figure}[!tb]
    \centering
    \includegraphics[width=0.48\linewidth]{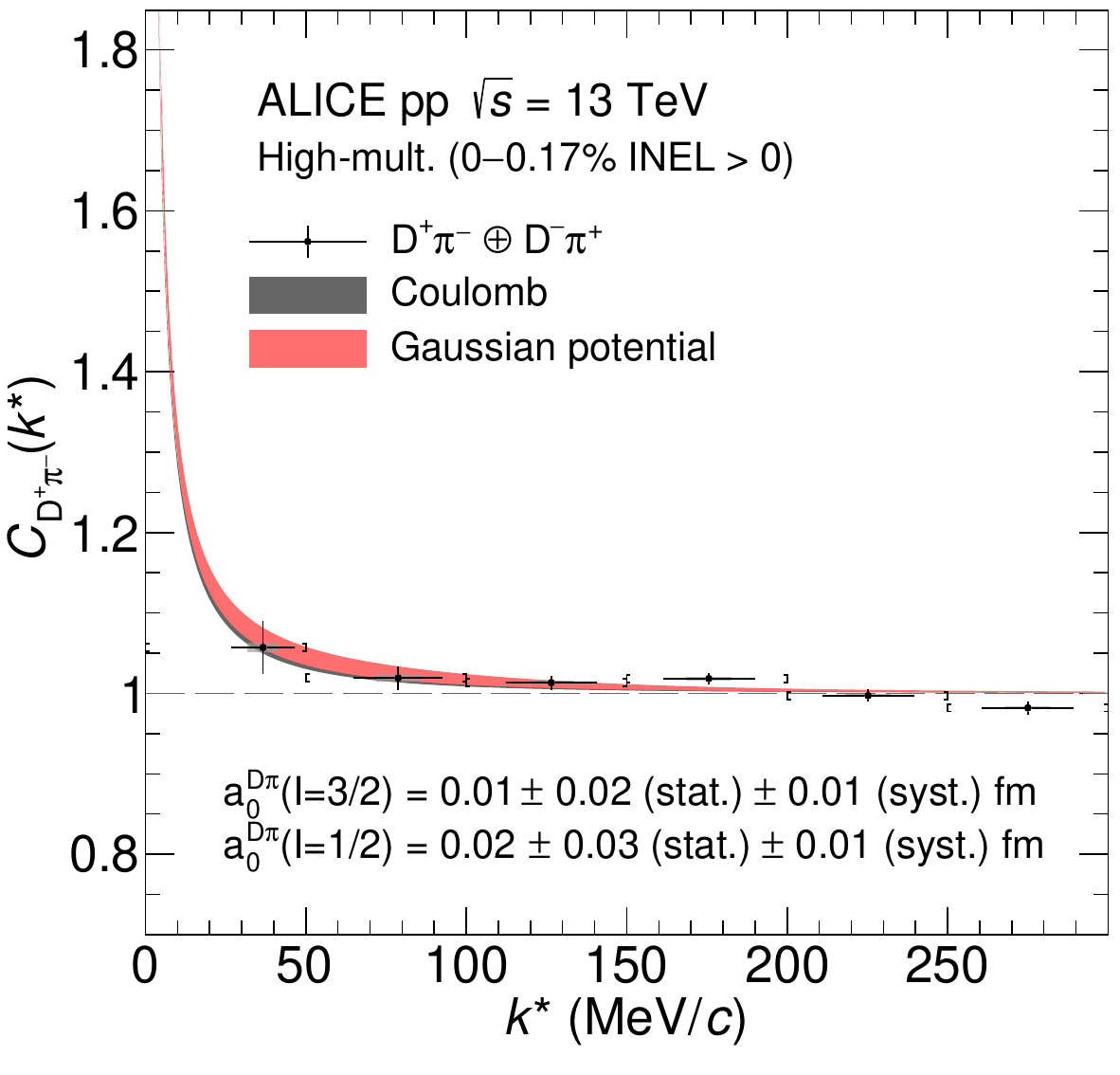}
    \includegraphics[width=0.48\linewidth]{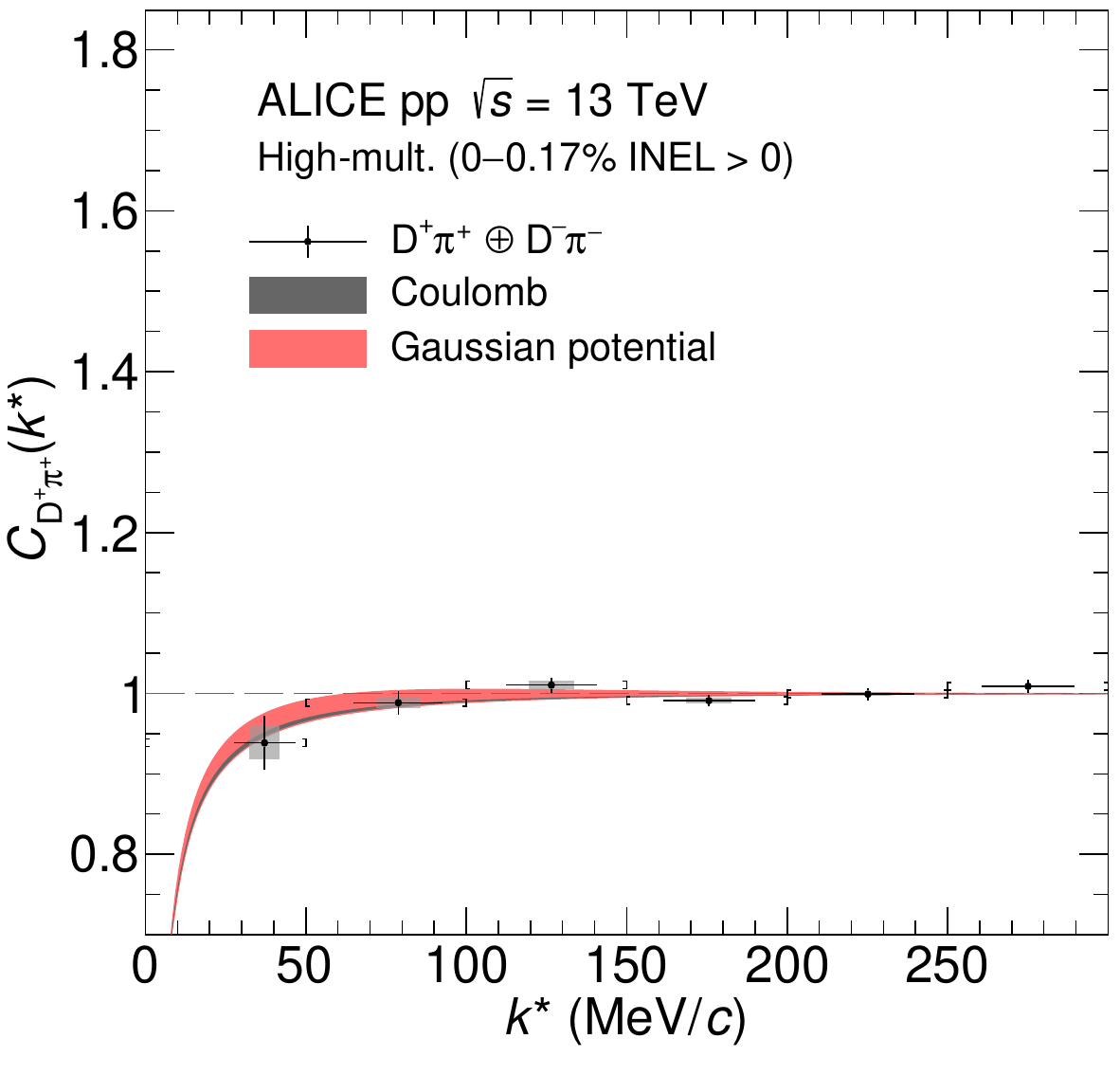}\\
    \includegraphics[width=0.48\linewidth]{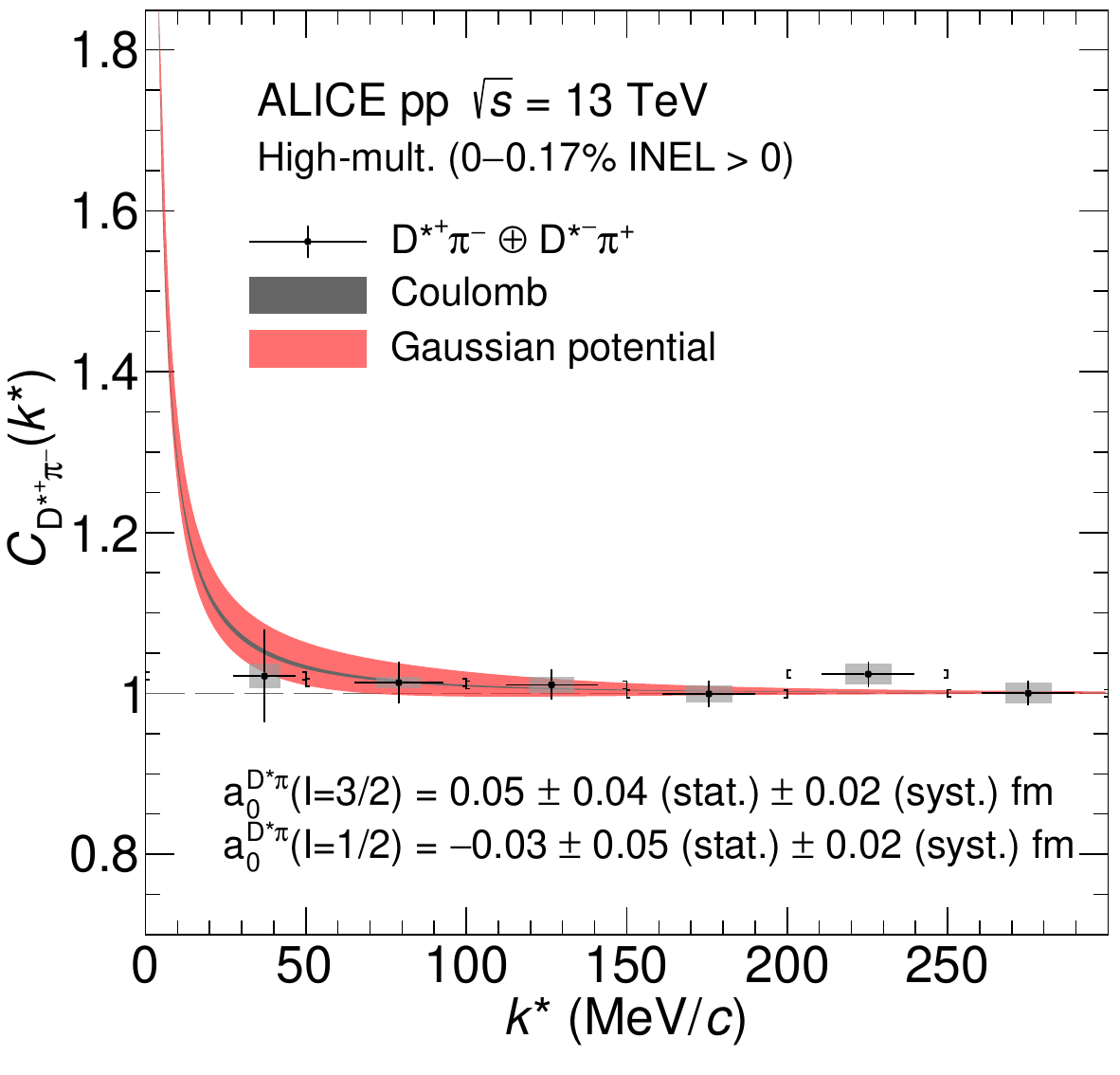}
    \includegraphics[width=0.48\linewidth]{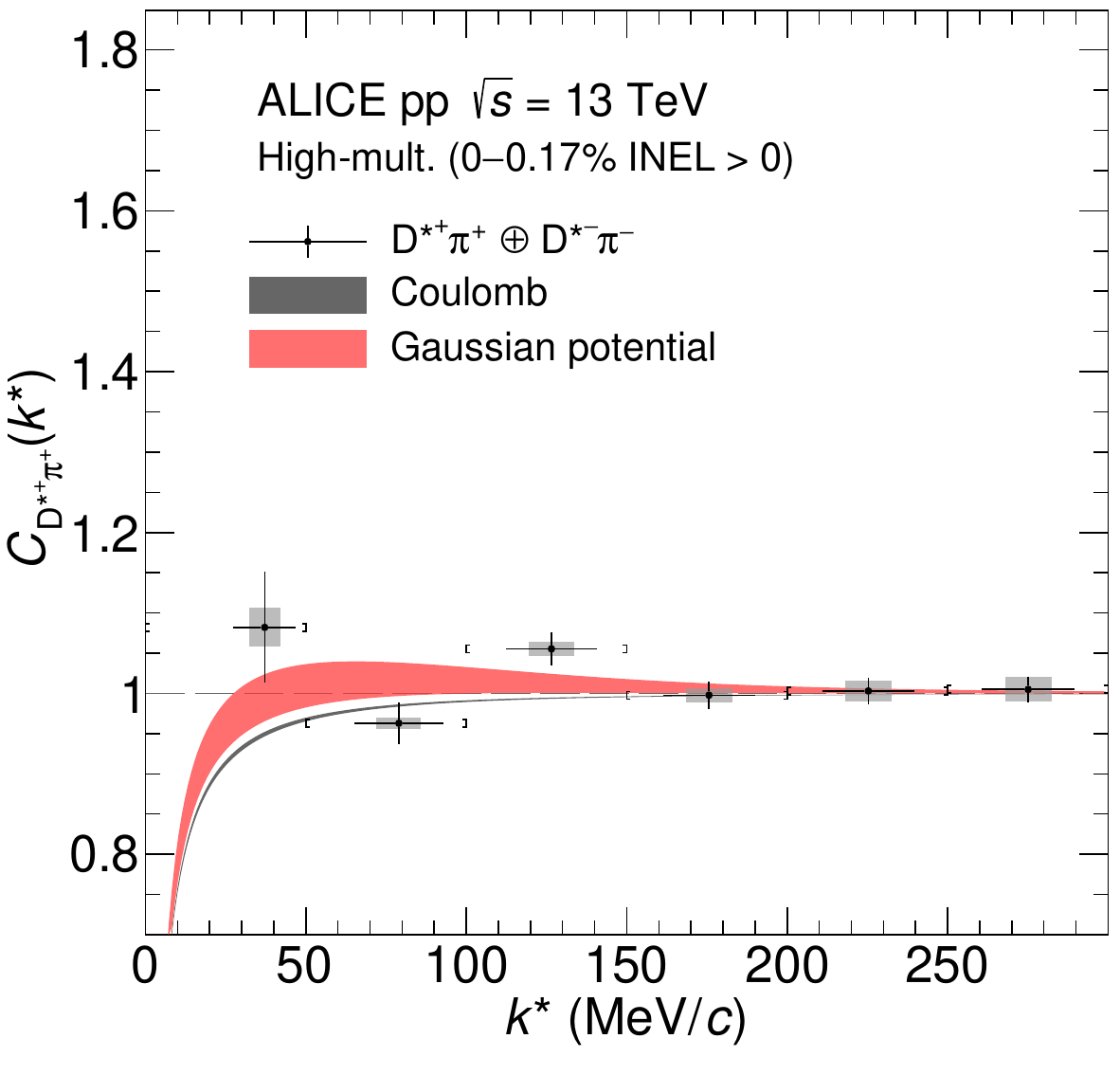}
    \caption{Comparison of the \DPi (first row) and \DstarPi (second row) genuine correlation functions of same- (left column) and opposite-charge (right column) combinations with the results of the $\chi^2$ minimization using a Gaussian potential to parameterize the strong interaction (red band). The width of the band corresponds to the total uncertainty. The gray curve represents the correlation functions assuming only interaction via the Coulomb force.}
    \label{fig:CFD}
\end{figure}

\begin{figure}[!tb]
    \centering
    \includegraphics[width=0.49\linewidth]{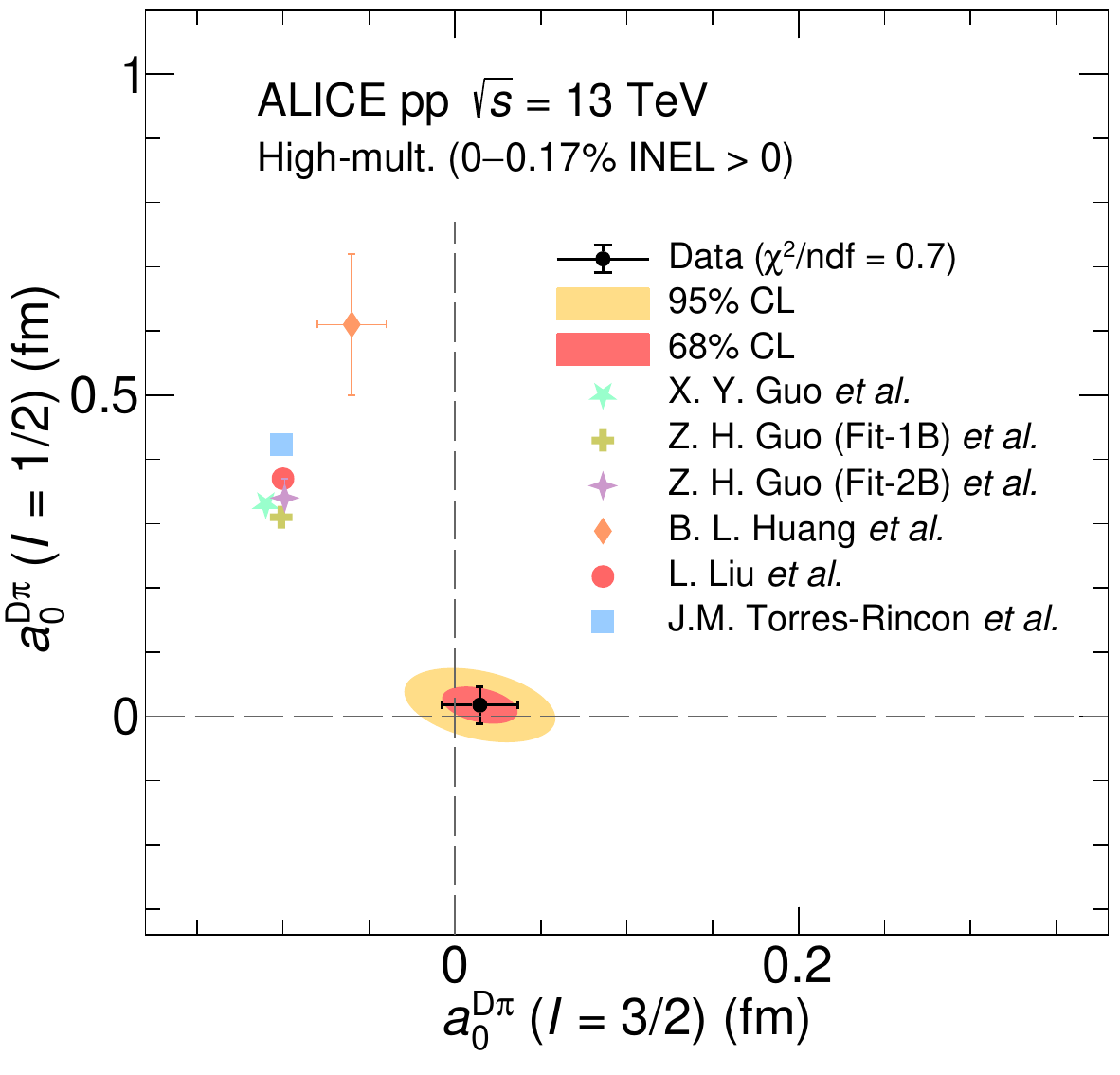}
    \includegraphics[width=0.49\linewidth]{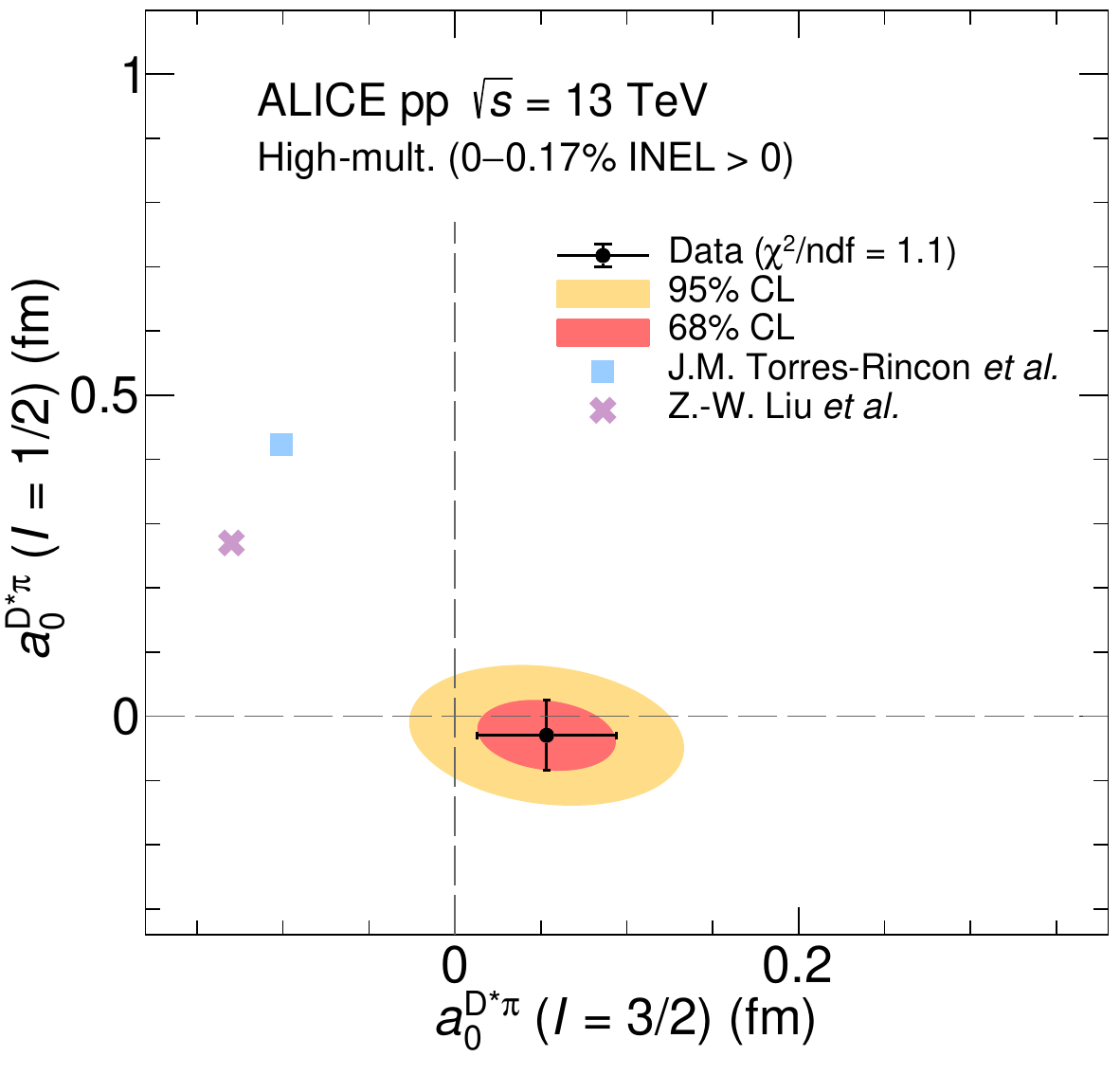}
    \caption{Scattering length of the \DPi (left) and \DstarPi (right) interaction, for the two isospin channels that characterize the systems. They are extracted from a simultaneous $\chi^2$ minimization to the experimental correlation functions. The red (orange) areas represent the resulting confidence intervals for a 68\% (95\%) probability. The dashed lines correspond to Coulomb interaction only, as the scattering lengths of the strong interaction vanish. As comparison, the available theoretical predictions~\cite{Liu:2012zya,Guo:2018kno,Guo:2018tjx,Huang:2021fdt,Torres-Rincon:2023qll, Liu:2011mi}, listed in Tables~\ref{tab:scatParamPi} and~\ref{tab:scatParamK} are shown as well.}
    \label{fig:CFDcorr}
\end{figure}

The theoretical correlation functions obtained from the different models of the strong interaction between charm and light-flavors mesons are compared to the measured genuine correlation functions in Fig.~\ref{fig:modelsdplus} and Fig.~\ref{fig:modelsdstar} for \Dplus and \Dstarp, respectively. The predictions for the Coulomb-only hypothesis (gray curves) are shown as a reference, as any deviation of the experimental data from it indicates the presence of strong FSI.
Additionally, the difference between the data and the calculations is quantified by the number of standard deviations $n_{\sigma}$ and is reported in the figure legends. Each $n_{\sigma}$ value is directly obtained from the $p$-value and reflects how well the specific model describes the data in the range of $\kstar<200~\MeVc$ by considering the total uncertainty of the data as well as the predictions.

Even though the current statistical precision is not sufficient to distinguish between the individual model predictions of the residual strong interaction involving kaons, no tension with theory is observed in most cases. The exception is \KDOC, where the larger $n_{\sigma}$ values are likely due to the fluctuation of the fourth data point. This is different for correlation functions involving pions. In the case of opposite-charge \DorDstarNoSignPi pairs the data are significantly lower than any of the model predictions and clearly favor the Coulomb-only hypothesis. For same-charge pairs, the deviation between data and models is much smaller, however, the Coulomb-only hypothesis is still favored.\\
In general, the correlation functions for all the analyzed particle systems can be adequately described by only considering the Coulomb interaction, indicating a shallow residual strong interaction between the \Dplus and \Dstarp mesons and light-flavor hadrons. A slight tension of $n_{\sigma}=2.62$ is observed for the \DstarPisc system, where the data points scatter around unity in the low \kstar region. However, as mentioned above, the Coulomb-only hypothesis is still favored over the calculations with residual strong interactions. In the case of \DKoc the $n_{\sigma}=2.72$ between the data and Coulomb-only hypotheses could be related to the fluctuating data point at $\kstar\sim180~\MeVc$. By only considering smaller $\kstar$ values, the $n_{\sigma}$ value reduces to 1.76, indicating that the Coulomb interaction sufficiently describes the measurement in the sensitive relative-momentum region.

The most precise correlation functions of this analysis, namely \DPisc and \DstarPisc, are employed to extract the scattering length $a_0$ of the strong interaction.
This is done by parameterizing the data using the same approach as for the theory predictions, which involves a Gaussian potential given by Eq.~\ref{eq:gaussianpot} with variable potential strength $V_0$ to model the strong interaction.
As the isospin $I=3/2$ state is shared among both charge combinations, the corresponding $V_0^{I=3/2}$ parameter is a common fit parameter of the two correlation functions. The potential strengths $V_0^{I=3/2}$ and $V_0^{I=1/2}$ are determined by a simultaneous $\chi^2$ minimization within $\kstar<250~\MeVc$.
Finally, the $I=1/2$ and $3/2$ scattering lengths are calculated by solving the Schr\"{o}dinger equation in the isospin basis. 

The scattering lengths extracted by the minimization are summarized in Table~\ref{tab:parres}. Figure~\ref{fig:CFD} shows the corresponding model correlation functions (red bands) as well as the fitted data. The width of the bands represents the total uncertainty obtained from the $\chi^2$ minimization. The $\chi^2/\mathrm{ndf}$ of the combined fit of the \DPi correlation functions is 0.7 within $\kstar<250~\MeVc$, while it is 1.0 in the case of \DstarPi.
The correlation between the scattering lengths for the isospin channels $I=1/2$ and $I=3/2$ extracted from the simultaneous fits are shown in Fig.~\ref{fig:CFDcorr}. The red (orange) areas represent the confidence intervals for a 68\% (95\%) probability. 
Notably, the scattering lengths governing the residual strong interaction between \Dplus and \Dstarp mesons with light-flavor mesons are found to be compatible with each other within the uncertainties.
This is understood in terms of heavy-quark spin symmetry, which states that, at leading order, the interaction of light-flavor mesons with pseudoscalar or vector charm mesons is the same.\\
The measured scattering lengths for the isospin $I=1/2$ channels are vanishing for both \Dplus and \Dstarp mesons. In the case of $I=3/2$, they are compatible with zero within uncertainties for the \DPi interaction, while they are positive with a significance of about $1.1\sigma$ for the \DstarPi interaction. The scattering lengths extracted from the data are further compared with the theoretical predictions reported in Tables~\ref{tab:scatParamPi} and~\ref{tab:scatParamK}. For the $I=1/2$ channel, the measurements are significantly different from the values predicted by theoretical models, which cover the range between about 0.3 and 0.6 fm. Depending on the model, $5-13\sigma$ are obtained for \DPi and $6-8\sigma$ for \DstarPi, taking into account the uncertainty of the data as well as the predictions. In the case of the $I=3/2$ channel, the measurements also show a tension with the theoretical predictions, it is, however, smaller than in the $I=1/2$ channel. 
In the \DPi case, a deviation of $2-5\sigma$ is found, depending on the model, while it is around $3-4\sigma$ for \DstarPi. 
A much larger source size could diminish this discrepancy, as it leads to a less pronounced correlation signal for a given interaction strength. However, there is no obvious motivation for assuming a breaking of the universal \mt scaling of the core radius~\cite{ALICE:2020ibs,KorwieserSource} in the case of correlation functions involving charm mesons. Especially, it is successfully used in the analysis of the experimental \pD correlation function~\cite{ALICE:2022enj}. 
In Ref.~\cite{Khemchandani:2023xup} the hidden gauge formalism, implementing unitarization in coupled channels, is used to study the molecular nature of the lowest-lying D$_1$ states (D$_1(2420)$ and D$_1(2430)$), as well as the scattering amplitudes of some of the members of the meson$-$baryon basis considered (\DstarPi, D$\uprho$) and the corresponding correlation functions. In order to better accommodate the D$_1(2430)$ within the experimental observations~\cite{Belle:2003nsh, LHCb:2015tsv}, a bare quark-model pole structure is added explicitly, whose parameters-dependence allows the authors to consider two plausible scenarios. The one denoted as Model B in their publication provides as a result a scattering length of $a_{\DstarPi}^{I=1/2}=0.1~$fm, which is a value much closer to the one obtained in the present work.
Alternatively, other complex structures, for example in higher partial waves, not taken into account by the theory models, could modify the predictions.

In summary, the measured correlation functions between charm mesons and light-flavor mesons are compatible with the predictions obtained with only Coulomb interaction, suggesting that the residual strong interaction between these pairs of particles is shallow. 
A significant discrepancy in the $I=1/2$ channel is found with respect to the predictions for the \DorDstarNoSignPi scattering lengths, which is much less pronounced in the $I=3/2$ channel. This discrepancy could be reconciled with the theory only in the case of a sizeable emitting source, which is not well motivated. The current precision of the \DorDstarNoSignK correlation functions does not allow for the discrimination between the available models and for a firm conclusion on the possible formation of bound states. Finally, the measured interactions suggest that the rescattering probability of charm mesons with light hadrons in the hadronic phase of the system produced in ultrarelativistic collisions is small. Even with values of scattering lengths predicted by theory calculations, which are larger than the measured ones reported in this article, a small impact on the D-meson final momenta is expected~\cite{He:2011yi}, given the duration of the hadronic phase of the system created in ultrarelativistic heavy-ion collisions of about $\Delta\tau_\mathrm{had} \approx 5-10~\fm/c$~\cite{He:2011yi,Abreu:2011ic,ALICE:2022wpn}.

\begin{table}[!tb]
\centering
\caption{The scattering lengths $a_0$ of the \DorDstarNoSignPi interaction, extracted from a $\chi^2$ minimization to the experimental genuine correlation function, using a Gaussian potential to parameterize the strong interaction. 
}
\renewcommand{\arraystretch}{1.2}
\scalebox{1}{
\begin{tabular}{l | r | r }
\multicolumn{1}{c|}{Pair} & \multicolumn{1}{c|}{$I$} & \multicolumn{1}{c}{$a_0$~[fm]}\\
\hline
\hline
\multirow{2}{*}{\DPi} & $3/2$  & $0.01  \pm 0.02~\mathrm{(stat.)} \pm 0.01~\mathrm{(syst.)}$\\
& $1/2$  & $0.02 \pm 0.03~\mathrm{(stat.)} \pm 0.01~\mathrm{(syst.)}$\\
\hline
\multirow{2}{*}{\DstarPi} &$3/2$ & $ 0.05 \pm 0.04~\mathrm{(stat.)} \pm 0.02~\mathrm{(syst.)}$\\
&$1/2$  & $-0.03 \pm 0.05~\mathrm{(stat.)} \pm 0.02~\mathrm{(syst.)}$\\
\hline
\hline
\end{tabular} 
}
\label{tab:parres}
\end{table}

\section{Conclusion}
\label{sec:conclusions}

The study of the residual strong interactions of \DorDstarp mesons with charged pions and kaons is performed for the first time, using high-multiplicity proton--proton collision data at \thirteen collected with the ALICE detector at the LHC.
The femtoscopy technique is used to test various theoretical models of the strong interaction by comparing the experimental correlation functions for the different particle pairs with the predictions by theory. As comparison also the Coulomb-only assumption is tested and, within the current uncertainties, all the measured correlation functions can be well described by it. For the same charge \DstarPi system, a slight tension with the Coulomb-only assumption of $n_{\sigma}=2.62$ is observed. Still, it describes the data better than the model including the strong interaction. 
A comparison of the \DK and \DstarK data to theoretical predictions does not lead to a clear result, as no preference among the different models of the strong interaction or Coulomb-only hypothesis is observed due to the limited statistical precision. In the case of \DPi interaction instead, the experimental data indicates that the theoretical models overestimate the scattering lengths, especially in the opposite-charge \DPi correlation function, where a strong discrepancy is found. In comparison, Coulomb-only predictions yield a better description of the data. The same can be observed for the correlation functions involving \Dstarp mesons. 

Among the experimental correlation functions studied in this work, the ones of the \DPi and \DstarPi systems are the most precise. Therefore, they are used to determine the scattering lengths of the strong interaction, which is modeled using a Gaussian potential. The scattering parameters are found to be small and compatible with zero. Especially, the disagreement between the scattering length of the isospin channel $I=1/2$, extracted from the data, and the theoretical predictions is found to be larger than $5\sigma$, challenging the current understanding of the residual strong interaction between D mesons and pions. 

These findings also provide important information for the interpretation of the measurements of D-meson production and angular anisotropy in heavy-ion collisions~\cite{ALICE:2020iug,ALICE:2021rxa}, since they suggest that the effect of the rescattering of \DorDstarp mesons with light hadrons during the hadronic phase of the system produced in such collisions is small.

The precision of these measurements will improve with the data taken during the LHC Run 3 data-taking period. In fact, the dataset collected by the ALICE Collaboration will benefit from various detector upgrades, which include an improved spatial resolution crucial for the reconstruction of heavy-flavor decay vertices, and a larger luminosity thanks to the higher readout rate achievable~\cite{ALICE:2023udb}. Furthermore, with such improvements, the momentum correlation functions of other particle pairs involving charm hadrons will also become accessible.


\newenvironment{acknowledgement}{\relax}{\relax}
\begin{acknowledgement}
\section*{Acknowledgements}

The ALICE Collaboration would like to thank all its engineers and technicians for their invaluable contributions to the construction of the experiment and the CERN accelerator teams for the outstanding performance of the LHC complex.
The ALICE Collaboration gratefully acknowledges the resources and support provided by all Grid centres and the Worldwide LHC Computing Grid (WLCG) collaboration.
The ALICE Collaboration acknowledges the following funding agencies for their support in building and running the ALICE detector:
A. I. Alikhanyan National Science Laboratory (Yerevan Physics Institute) Foundation (ANSL), State Committee of Science and World Federation of Scientists (WFS), Armenia;
Austrian Academy of Sciences, Austrian Science Fund (FWF): [M 2467-N36] and Nationalstiftung f\"{u}r Forschung, Technologie und Entwicklung, Austria;
Ministry of Communications and High Technologies, National Nuclear Research Center, Azerbaijan;
Conselho Nacional de Desenvolvimento Cient\'{\i}fico e Tecnol\'{o}gico (CNPq), Financiadora de Estudos e Projetos (Finep), Funda\c{c}\~{a}o de Amparo \`{a} Pesquisa do Estado de S\~{a}o Paulo (FAPESP) and Universidade Federal do Rio Grande do Sul (UFRGS), Brazil;
Bulgarian Ministry of Education and Science, within the National Roadmap for Research Infrastructures 2020-2027 (object CERN), Bulgaria;
Ministry of Education of China (MOEC) , Ministry of Science \& Technology of China (MSTC) and National Natural Science Foundation of China (NSFC), China;
Ministry of Science and Education and Croatian Science Foundation, Croatia;
Centro de Aplicaciones Tecnol\'{o}gicas y Desarrollo Nuclear (CEADEN), Cubaenerg\'{\i}a, Cuba;
Ministry of Education, Youth and Sports of the Czech Republic, Czech Republic;
The Danish Council for Independent Research | Natural Sciences, the VILLUM FONDEN and Danish National Research Foundation (DNRF), Denmark;
Helsinki Institute of Physics (HIP), Finland;
Commissariat \`{a} l'Energie Atomique (CEA) and Institut National de Physique Nucl\'{e}aire et de Physique des Particules (IN2P3) and Centre National de la Recherche Scientifique (CNRS), France;
Bundesministerium f\"{u}r Bildung und Forschung (BMBF) and GSI Helmholtzzentrum f\"{u}r Schwerionenforschung GmbH, Germany;
General Secretariat for Research and Technology, Ministry of Education, Research and Religions, Greece;
National Research, Development and Innovation Office, Hungary;
Department of Atomic Energy Government of India (DAE), Department of Science and Technology, Government of India (DST), University Grants Commission, Government of India (UGC) and Council of Scientific and Industrial Research (CSIR), India;
National Research and Innovation Agency - BRIN, Indonesia;
Istituto Nazionale di Fisica Nucleare (INFN), Italy;
Japanese Ministry of Education, Culture, Sports, Science and Technology (MEXT) and Japan Society for the Promotion of Science (JSPS) KAKENHI, Japan;
Consejo Nacional de Ciencia (CONACYT) y Tecnolog\'{i}a, through Fondo de Cooperaci\'{o}n Internacional en Ciencia y Tecnolog\'{i}a (FONCICYT) and Direcci\'{o}n General de Asuntos del Personal Academico (DGAPA), Mexico;
Nederlandse Organisatie voor Wetenschappelijk Onderzoek (NWO), Netherlands;
The Research Council of Norway, Norway;
Commission on Science and Technology for Sustainable Development in the South (COMSATS), Pakistan;
Pontificia Universidad Cat\'{o}lica del Per\'{u}, Peru;
Ministry of Education and Science, National Science Centre and WUT ID-UB, Poland;
Korea Institute of Science and Technology Information and National Research Foundation of Korea (NRF), Republic of Korea;
Ministry of Education and Scientific Research, Institute of Atomic Physics, Ministry of Research and Innovation and Institute of Atomic Physics and Universitatea Nationala de Stiinta si Tehnologie Politehnica Bucuresti, Romania;
Ministry of Education, Science, Research and Sport of the Slovak Republic, Slovakia;
National Research Foundation of South Africa, South Africa;
Swedish Research Council (VR) and Knut \& Alice Wallenberg Foundation (KAW), Sweden;
European Organization for Nuclear Research, Switzerland;
Suranaree University of Technology (SUT), National Science and Technology Development Agency (NSTDA) and National Science, Research and Innovation Fund (NSRF via PMU-B B05F650021), Thailand;
Turkish Energy, Nuclear and Mineral Research Agency (TENMAK), Turkey;
National Academy of  Sciences of Ukraine, Ukraine;
Science and Technology Facilities Council (STFC), United Kingdom;
National Science Foundation of the United States of America (NSF) and United States Department of Energy, Office of Nuclear Physics (DOE NP), United States of America.
In addition, individual groups or members have received support from:
Czech Science Foundation (grant no. 23-07499S), Czech Republic;
European Research Council (grant no. 950692), European Union;
ICSC - Centro Nazionale di Ricerca in High Performance Computing, Big Data and Quantum Computing, European Union - NextGenerationEU;
Academy of Finland (Center of Excellence in Quark Matter) (grant nos. 346327, 346328), Finland.

\end{acknowledgement}

\bibliographystyle{utphys}   
\bibliography{bibliography}

\newpage
\appendix

%
%

\section{The ALICE Collaboration}
\label{app:collab}
\begin{flushleft} 
\small

S.~Acharya\,\orcidlink{0000-0002-9213-5329}\,$^{\rm 129}$, 
D.~Adamov\'{a}\,\orcidlink{0000-0002-0504-7428}\,$^{\rm 87}$, 
G.~Aglieri Rinella\,\orcidlink{0000-0002-9611-3696}\,$^{\rm 33}$, 
L.~Aglietta$^{\rm 25}$, 
M.~Agnello\,\orcidlink{0000-0002-0760-5075}\,$^{\rm 30}$, 
N.~Agrawal\,\orcidlink{0000-0003-0348-9836}\,$^{\rm 26}$, 
Z.~Ahammed\,\orcidlink{0000-0001-5241-7412}\,$^{\rm 137}$, 
S.~Ahmad\,\orcidlink{0000-0003-0497-5705}\,$^{\rm 16}$, 
S.U.~Ahn\,\orcidlink{0000-0001-8847-489X}\,$^{\rm 72}$, 
I.~Ahuja\,\orcidlink{0000-0002-4417-1392}\,$^{\rm 38}$
% ...

\section*{Affiliation Notes}

$^{\rm I}$ Deceased\\
$^{\rm II}$ Also at: Max-Planck-Institut fur Physik, Munich, Germany\\
$^{\rm III}$ Also at: Italian National Agency for New Technologies, Energy and Sustainable Economic Development (ENEA), Bologna, Italy\\
$^{\rm IV}$ Also at: Dipartimento DET del Politecnico di Torino, Turin, Italy\\
$^{\rm V}$ Also at: Helmholtz Institut f\"ur Strahlen- und Kernphysik and Bethe Center for Theoretical Physics, Universit\"at Bonn, Bonn, Germany\\
$^{\rm VI}$ Also at: Yildiz Technical University, Istanbul, T\"{u}rkiye\\
$^{\rm VII}$ Also at: Department of Applied Physics, Aligarh Muslim University, Aligarh, India\\
$^{\rm VIII}$ Also at: Institute of Theoretical Physics, University of Wroclaw, Poland\\
$^{\rm IX}$ Also at: An institution covered by a cooperation agreement with CERN\\

\section*{Collaboration Institutes}

$^{1}$ A.I. Alikhanyan National Science Laboratory (Yerevan Physics Institute) Foundation, Yerevan, Armenia\\
$^{2}$ AGH University of Krakow, Cracow, Poland\\
$^{3}$ Bogolyubov Institute for Theoretical Physics, National Academy of Sciences of Ukraine, Kiev, Ukraine\\
$^{4}$ Bose Institute, Department of Physics  and Centre for Astroparticle Physics and Space Science (CAPSS), Kolkata, India\\
$^{5}$ California Polytechnic State University, San Luis Obispo, California, United States\\
$^{6}$ Central China Normal University, Wuhan, China\\
$^{7}$ Centro de Aplicaciones Tecnol\'{o}gicas y Desarrollo Nuclear (CEADEN), Havana, Cuba\\
$^{8}$ Centro de Investigaci\'{o}n y de Estudios Avanzados (CINVESTAV), Mexico City and M\'{e}rida, Mexico\\
$^{9}$ Chicago State University, Chicago, Illinois, United States\\
$^{10}$ China Institute of Atomic Energy, Beijing, China\\
$^{11}$ China University of Geosciences, Wuhan, China\\
$^{12}$ Chungbuk National University, Cheongju, Republic of Korea\\
$^{13}$ Comenius University Bratislava, Faculty of Mathematics, Physics and Informatics, Bratislava, Slovak Republic\\
$^{14}$ COMSATS University Islamabad, Islamabad, Pakistan\\
$^{15}$ Creighton University, Omaha, Nebraska, United States\\
$^{16}$ Department of Physics, Aligarh Muslim University, Aligarh, India\\
$^{17}$ Department of Physics, Pusan National University, Pusan, Republic of Korea\\
$^{18}$ Department of Physics, Sejong University, Seoul, Republic of Korea\\
$^{19}$ Department of Physics, University of California, Berkeley, California, United States\\
$^{20}$ Department of Physics, University of Oslo, Oslo, Norway\\
$^{21}$ Department of Physics and Technology, University of Bergen, Bergen, Norway\\
$^{22}$ Dipartimento di Fisica, Universit\`{a} di Pavia, Pavia, Italy\\
$^{23}$ Dipartimento di Fisica dell'Universit\`{a} and Sezione INFN, Cagliari, Italy\\
$^{24}$ Dipartimento di Fisica dell'Universit\`{a} and Sezione INFN, Trieste, Italy\\
$^{25}$ Dipartimento di Fisica dell'Universit\`{a} and Sezione INFN, Turin, Italy\\
$^{26}$ Dipartimento di Fisica e Astronomia dell'Universit\`{a} and Sezione INFN, Bologna, Italy\\
$^{27}$ Dipartimento di Fisica e Astronomia dell'Universit\`{a} and Sezione INFN, Catania, Italy\\
$^{28}$ Dipartimento di Fisica e Astronomia dell'Universit\`{a} and Sezione INFN, Padova, Italy\\
$^{29}$ Dipartimento di Fisica `E.R.~Caianiello' dell'Universit\`{a} and Gruppo Collegato INFN, Salerno, Italy\\
$^{30}$ Dipartimento DISAT del Politecnico and Sezione INFN, Turin, Italy\\
$^{31}$ Dipartimento di Scienze MIFT, Universit\`{a} di Messina, Messina, Italy\\
$^{32}$ Dipartimento Interateneo di Fisica `M.~Merlin' and Sezione INFN, Bari, Italy\\
$^{33}$ European Organization for Nuclear Research (CERN), Geneva, Switzerland\\
$^{34}$ Faculty of Electrical Engineering, Mechanical Engineering and Naval Architecture, University of Split, Split, Croatia\\
$^{35}$ Faculty of Engineering and Science, Western Norway University of Applied Sciences, Bergen, Norway\\
$^{36}$ Faculty of Nuclear Sciences and Physical Engineering, Czech Technical University in Prague, Prague, Czech Republic\\
$^{37}$ Faculty of Physics, Sofia University, Sofia, Bulgaria\\
$^{38}$ Faculty of Science, P.J.~\v{S}af\'{a}rik University, Ko\v{s}ice, Slovak Republic\\
$^{39}$ Frankfurt Institute for Advanced Studies, Johann Wolfgang Goethe-Universit\"{a}t Frankfurt, Frankfurt, Germany\\
$^{40}$ Fudan University, Shanghai, China\\
$^{41}$ Gangneung-Wonju National University, Gangneung, Republic of Korea\\
$^{42}$ Gauhati University, Department of Physics, Guwahati, India\\
$^{43}$ Helmholtz-Institut f\"{u}r Strahlen- und Kernphysik, Rheinische Friedrich-Wilhelms-Universit\"{a}t Bonn, Bonn, Germany\\
$^{44}$ Helsinki Institute of Physics (HIP), Helsinki, Finland\\
$^{45}$ High Energy Physics Group,  Universidad Aut\'{o}noma de Puebla, Puebla, Mexico\\
$^{46}$ Horia Hulubei National Institute of Physics and Nuclear Engineering, Bucharest, Romania\\
$^{47}$ HUN-REN Wigner Research Centre for Physics, Budapest, Hungary\\
$^{48}$ Indian Institute of Technology Bombay (IIT), Mumbai, India\\
$^{49}$ Indian Institute of Technology Indore, Indore, India\\
$^{50}$ INFN, Laboratori Nazionali di Frascati, Frascati, Italy\\
$^{51}$ INFN, Sezione di Bari, Bari, Italy\\
$^{52}$ INFN, Sezione di Bologna, Bologna, Italy\\
$^{53}$ INFN, Sezione di Cagliari, Cagliari, Italy\\
$^{54}$ INFN, Sezione di Catania, Catania, Italy\\
$^{55}$ INFN, Sezione di Padova, Padova, Italy\\
$^{56}$ INFN, Sezione di Pavia, Pavia, Italy\\
$^{57}$ INFN, Sezione di Torino, Turin, Italy\\
$^{58}$ INFN, Sezione di Trieste, Trieste, Italy\\
$^{59}$ Inha University, Incheon, Republic of Korea\\
$^{60}$ Institute for Gravitational and Subatomic Physics (GRASP), Utrecht University/Nikhef, Utrecht, Netherlands\\
$^{61}$ Institute of Experimental Physics, Slovak Academy of Sciences, Ko\v{s}ice, Slovak Republic\\
$^{62}$ Institute of Physics, Homi Bhabha National Institute, Bhubaneswar, India\\
$^{63}$ Institute of Physics of the Czech Academy of Sciences, Prague, Czech Republic\\
$^{64}$ Institute of Space Science (ISS), Bucharest, Romania\\
$^{65}$ Institut f\"{u}r Kernphysik, Johann Wolfgang Goethe-Universit\"{a}t Frankfurt, Frankfurt, Germany\\
$^{66}$ Instituto de Ciencias Nucleares, Universidad Nacional Aut\'{o}noma de M\'{e}xico, Mexico City, Mexico\\
$^{67}$ Instituto de F\'{i}sica, Universidade Federal do Rio Grande do Sul (UFRGS), Porto Alegre, Brazil\\
$^{68}$ Instituto de F\'{\i}sica, Universidad Nacional Aut\'{o}noma de M\'{e}xico, Mexico City, Mexico\\
$^{69}$ iThemba LABS, National Research Foundation, Somerset West, South Africa\\
$^{70}$ Jeonbuk National University, Jeonju, Republic of Korea\\
$^{71}$ Johann-Wolfgang-Goethe Universit\"{a}t Frankfurt Institut f\"{u}r Informatik, Fachbereich Informatik und Mathematik, Frankfurt, Germany\\
$^{72}$ Korea Institute of Science and Technology Information, Daejeon, Republic of Korea\\
$^{73}$ KTO Karatay University, Konya, Turkey\\
$^{74}$ Laboratoire de Physique Subatomique et de Cosmologie, Universit\'{e} Grenoble-Alpes, CNRS-IN2P3, Grenoble, France\\
$^{75}$ Lawrence Berkeley National Laboratory, Berkeley, California, United States\\
$^{76}$ Lund University Department of Physics, Division of Particle Physics, Lund, Sweden\\
$^{77}$ Nagasaki Institute of Applied Science, Nagasaki, Japan\\
$^{78}$ Nara Women{'}s University (NWU), Nara, Japan\\
$^{79}$ National and Kapodistrian University of Athens, School of Science, Department of Physics , Athens, Greece\\
$^{80}$ National Centre for Nuclear Research, Warsaw, Poland\\
$^{81}$ National Institute of Science Education and Research, Homi Bhabha National Institute, Jatni, India\\
$^{82}$ National Nuclear Research Center, Baku, Azerbaijan\\
$^{83}$ National Research and Innovation Agency - BRIN, Jakarta, Indonesia\\
$^{84}$ Niels Bohr Institute, University of Copenhagen, Copenhagen, Denmark\\
$^{85}$ Nikhef, National institute for subatomic physics, Amsterdam, Netherlands\\
$^{86}$ Nuclear Physics Group, STFC Daresbury Laboratory, Daresbury, United Kingdom\\
$^{87}$ Nuclear Physics Institute of the Czech Academy of Sciences, Husinec-\v{R}e\v{z}, Czech Republic\\
$^{88}$ Oak Ridge National Laboratory, Oak Ridge, Tennessee, United States\\
$^{89}$ Ohio State University, Columbus, Ohio, United States\\
$^{90}$ Physics department, Faculty of science, University of Zagreb, Zagreb, Croatia\\
$^{91}$ Physics Department, Panjab University, Chandigarh, India\\
$^{92}$ Physics Department, University of Jammu, Jammu, India\\
$^{93}$ Physics Program and International Institute for Sustainability with Knotted Chiral Meta Matter (SKCM2), Hiroshima University, Hiroshima, Japan\\
$^{94}$ Physikalisches Institut, Eberhard-Karls-Universit\"{a}t T\"{u}bingen, T\"{u}bingen, Germany\\
$^{95}$ Physikalisches Institut, Ruprecht-Karls-Universit\"{a}t Heidelberg, Heidelberg, Germany\\
$^{96}$ Physik Department, Technische Universit\"{a}t M\"{u}nchen, Munich, Germany\\
$^{97}$ Politecnico di Bari and Sezione INFN, Bari, Italy\\
$^{98}$ Research Division and ExtreMe Matter Institute EMMI, GSI Helmholtzzentrum f\"ur Schwerionenforschung GmbH, Darmstadt, Germany\\
$^{99}$ RIKEN iTHEMS, Wako, Japan\\
$^{100}$ Saga University, Saga, Japan\\
$^{101}$ Saha Institute of Nuclear Physics, Homi Bhabha National Institute, Kolkata, India\\
$^{102}$ School of Physics and Astronomy, University of Birmingham, Birmingham, United Kingdom\\
$^{103}$ Secci\'{o}n F\'{\i}sica, Departamento de Ciencias, Pontificia Universidad Cat\'{o}lica del Per\'{u}, Lima, Peru\\
$^{104}$ Stefan Meyer Institut f\"{u}r Subatomare Physik (SMI), Vienna, Austria\\
$^{105}$ SUBATECH, IMT Atlantique, Nantes Universit\'{e}, CNRS-IN2P3, Nantes, France\\
$^{106}$ Sungkyunkwan University, Suwon City, Republic of Korea\\
$^{107}$ Suranaree University of Technology, Nakhon Ratchasima, Thailand\\
$^{108}$ Technical University of Ko\v{s}ice, Ko\v{s}ice, Slovak Republic\\
$^{109}$ The Henryk Niewodniczanski Institute of Nuclear Physics, Polish Academy of Sciences, Cracow, Poland\\
$^{110}$ The University of Texas at Austin, Austin, Texas, United States\\
$^{111}$ Universidad Aut\'{o}noma de Sinaloa, Culiac\'{a}n, Mexico\\
$^{112}$ Universidade de S\~{a}o Paulo (USP), S\~{a}o Paulo, Brazil\\
$^{113}$ Universidade Estadual de Campinas (UNICAMP), Campinas, Brazil\\
$^{114}$ Universidade Federal do ABC, Santo Andre, Brazil\\
$^{115}$ Universitatea Nationala de Stiinta si Tehnologie Politehnica Bucuresti, Bucharest, Romania\\
$^{116}$ University of Cape Town, Cape Town, South Africa\\
$^{117}$ University of Derby, Derby, United Kingdom\\
$^{118}$ University of Houston, Houston, Texas, United States\\
$^{119}$ University of Jyv\"{a}skyl\"{a}, Jyv\"{a}skyl\"{a}, Finland\\
$^{120}$ University of Kansas, Lawrence, Kansas, United States\\
$^{121}$ University of Liverpool, Liverpool, United Kingdom\\
$^{122}$ University of Science and Technology of China, Hefei, China\\
$^{123}$ University of South-Eastern Norway, Kongsberg, Norway\\
$^{124}$ University of Tennessee, Knoxville, Tennessee, United States\\
$^{125}$ University of the Witwatersrand, Johannesburg, South Africa\\
$^{126}$ University of Tokyo, Tokyo, Japan\\
$^{127}$ University of Tsukuba, Tsukuba, Japan\\
$^{128}$ Universit\"{a}t M\"{u}nster, Institut f\"{u}r Kernphysik, M\"{u}nster, Germany\\
$^{129}$ Universit\'{e} Clermont Auvergne, CNRS/IN2P3, LPC, Clermont-Ferrand, France\\
$^{130}$ Universit\'{e} de Lyon, CNRS/IN2P3, Institut de Physique des 2 Infinis de Lyon, Lyon, France\\
$^{131}$ Universit\'{e} de Strasbourg, CNRS, IPHC UMR 7178, F-67000 Strasbourg, France, Strasbourg, France\\
$^{132}$ Universit\'{e} Paris-Saclay, Centre d'Etudes de Saclay (CEA), IRFU, D\'{e}partment de Physique Nucl\'{e}aire (DPhN), Saclay, France\\
$^{133}$ Universit\'{e}  Paris-Saclay, CNRS/IN2P3, IJCLab, Orsay, France\\
$^{134}$ Universit\`{a} degli Studi di Foggia, Foggia, Italy\\
$^{135}$ Universit\`{a} del Piemonte Orientale, Vercelli, Italy\\
$^{136}$ Universit\`{a} di Brescia, Brescia, Italy\\
$^{137}$ Variable Energy Cyclotron Centre, Homi Bhabha National Institute, Kolkata, India\\
$^{138}$ Warsaw University of Technology, Warsaw, Poland\\
$^{139}$ Wayne State University, Detroit, Michigan, United States\\
$^{140}$ Yale University, New Haven, Connecticut, United States\\
$^{141}$ Yonsei University, Seoul, Republic of Korea\\
$^{142}$  Zentrum  f\"{u}r Technologie und Transfer (ZTT), Worms, Germany\\
$^{143}$ Affiliated with an institute covered by a cooperation agreement with CERN\\
$^{144}$ Affiliated with an international laboratory covered by a cooperation agreement with CERN.\\

\end{flushleft} 

\end{document}